\documentclass[onecolumn]{emulateapj}
\usepackage{epsfig}
\usepackage{amsmath}    

\def\etal{{et al. }}

\def\Dwa{$\,$\uppercase\expandafter{\romannumeral5}$\,$}

\def\sles{\lower2pt\hbox{$\buildrel {\scriptstyle <}
   \over {\scriptstyle\sim}$}}

\def\sgreat{\lower2pt\hbox{$\buildrel {\scriptstyle >}
   \over {\scriptstyle\sim}$}}
\def\sharpnull#1{}

\begin{document}
\def \jcp {J. of Comp. Phys.}
\def \m {$ M_\odot$~}
\def \l {$ L_\odot$}
\def \ro {g/cm$^{3}$}
\def \f {{\nu}}
\def \pos {{\mathbf r}}
\def \vel {{\mathbf v}}
\def \dir {{\mathbf \Omega}}
\def \dphidt {{\partial \Phi \over {\partial t}}}
\def \bn {{\mathbf n}}
\def \bB {{\mathbf B}}
\def \bBovr {(\frac{\mathbf B}{\rho})}
\def \bv {{\mathbf v}}
\def \Bf {{B_{\phi}}}
\def \vf {{v_{\phi}}}
\def \half {\frac{1}{2}} 
\def \ovr  {\frac{1}{r}}
\def \dpdr {{\partial p \over {\partial r}}}
\def \dpdz {{\partial p \over {\partial z}}}

\newcommand{\der}[2]{\frac{D#1}{D#2}}
\newcommand{\Dert}[1]{\frac{d#1}{dt}}
\newcommand{\dpar}[2]{\frac{\partial#1}{\partial#2}}
\newcommand{\dpart}[1]{\frac{\partial#1}{\partial t}}
\newcommand{\E}[2]{$#1\times 10^{#2}$}

\slugcomment{\bf}
\slugcomment{Accepted to Ap.J.}

\title{A Two-Dimensional MagnetoHydrodynamics Scheme for General Unstructured Grids}

\author{Eli Livne\altaffilmark{1}, Luc Dessart\altaffilmark{2}, 
         Adam Burrows\altaffilmark{2}, Casey A. Meakin\altaffilmark{2}}

\altaffiltext{1}{Racah Institute of Physics, The Hebrew University, Jerusalem,  Israel}
\altaffiltext{2}{Department of Astronomy and Steward Observatory, 
                 The University of Arizona, Tucson, AZ \ 85721}
\begin{abstract}

We report a new finite-difference scheme for two-dimensional magnetohydrodynamics (MHD)
simulations, with and without rotation, in unstructured grids with quadrilateral cells. 
The new scheme is implemented within the code VULCAN/2D, which already includes 
radiation-hydrodynamics in various approximations and can be used with arbitrarily
moving meshes (ALE). The MHD scheme, which consists of cell-centered magnetic field variables,
preserves the nodal finite difference representation of $div(\bB)$ by construction, and therefore
any initially divergence-free field remains divergence-free through the simulation.
In this paper, we describe the new scheme in detail and present comparisons of VULCAN/2D 
results with those of the code ZEUS/2D for several one-dimensional and two-dimensional test problems. 
The code now enables two-dimensional simulations of the collapse and explosion 
of the rotating, magnetic cores of massive stars. Moreover, it can be used to simulate the very 
wide variety of astrophysical problems for which multi-D radiation-magnetohydrodynamics 
(RMHD) is relevant.

\end{abstract}


\section{Introduction}
\label{intro}

Magnetic fields play a role, sometimes pivotal, in the evolution
and dynamics of astrophysical objects.  These include, but are not limited to,
molecular clouds, protostars, stellar and planetary dynamos, pulsars, jets from active galactic
nuclei, the solar wind, solar flares, the Earth's and Sun's magnetosheaths, 
and, importantly, accretion disks of all kinds.  To investigate these environments
theoretically has often required multi-dimensional simulation tools that in the past have incorporated
the relevant physics only to varying degrees.  The {\it ideal} MHD (magnetohydrodynamic) 
approximation, in which it is assumed the electrical conductivity of the fluid
is ``infinite" and, therefore, that the magnetic flux is frozen in the matter,
is a very good one in many environments in the Universe and obviates the necessity
to solve separate Boltzmann equations with microscopic couplings for the various
charged components.  

Recently, core-collapse supernova explosions and gamma-ray bursts have been 
added to the list of astrophysical sites in which magnetic 
fields might play an important role. In this paper, we describe 
a new 2.5-dimensional MHD algorithm (now incorporated into the 
radiation hydrodynamic code VULCAN/2D) that maintains the divergence-free
character of the B-field to machine precision and applies to general structured 
and unstructured grids.  We were motivated in developing this new capability
by the intriguing possibility that magnetic fields might play a dynamical 
role in the mechanism of gamma-ray bursts and, possibly, in the explosions 
of a subset of core-collapse supernovae, as well as by an interest in the 
origin of pulsar and magnetar B-fields. Nevertheless, the new code 
can be used for a much wider set of problems, since along with rotation 
and magnetic fields in the {\it ideal} MHD approximation it also incorporates multi-group 
transport (in both the multi-angle S$_n$ and flux-limited diffusion 
approximations) and a general gravity solver. 

The possibility of hydromagnetic driving of supernova 
explosions was first explored quantitatively by LeBlanc \& Wilson (1970).
Using numerical simulations that were then the state-of-the-art, they showed that the combination
of rapid rotation and strong magnetic fields in the pre-collapse core could lead to the formation of 
bipolar jets in the post-bounce configuration. In that picture, magnetic fields 
convert rotational energy into the jet energy via the pinch effect. The old
simulations of LeBlanc \& Wilson, however, suffered from crude spatial 
resolution and did not incorporate important developments in the physics 
of nuclear matter and in the neutrino-matter interaction. They employed 
gray transport and obtained rather weak explosions near $\sim$10$^{50}$ ergs.
After this pioneering work, with few exceptions (Bisnovatyi-Kogan et al. 
1976; Meier et al. 1976; Symbalisty 1984), the possible role of 
magnetic fields in powering or enabling supernova explosions took 
a back seat to other promising mechanisms. However, recently the subject of
hydromagnetic interactions during core collapse has been revived by 
a number of investigators (Akiyama et al. 2003; Ardeljan et al. 2005; Yamada 
\& Sawai 2004; Kotake et al. 2004; Ohnishi, 
Kotake, \& Yamada 2005; Akiyama \& Wheeler 2005; Proga 2005; 
Wilson, Mathews, \& Dalhed 2005; Thompson, Quataert, \& Burrows 2005; 
Moiseenko et al. 2006; Uzdensky \& MacFadyen 2006ab; Masada, Sano, \& Shibata 2006),
who have investigated the potential of magnetic stresses, magnetic buoyancy, magnetic collimation,
magnetic dissipation, and the magnetorotational instability (MRI, Balbus \& Hawley 1991; 
 Akiyama et al. 2003; Pessah \& Psaltis 2005; Pessah, Chan \& Psaltis 2006; 
 Obergaulinger, Aloy, \& M\"uller 2006; Etienne, Liu, \& Shapiro 2006)
 to power explosion.  In addition, the possibility that magnetic fields generated
during black hole or neutron star formation might form energetic jets in the context 
of gamma-ray bursts, X-ray flashes, collapsars (MacFadyen \& Woosley 1999), and hypernovae 
is receiving increasing attention (e.g., Mizuno et al. 2004). Furthermore, the potential role 
of magnetic winds in powering explosions, or secondary explosions, and in
spinning down the nascent neutron star or magnetar is now being actively studied 
(Thompson,~Chang,~\&~Quataert 2004; Bucciantini et al. 2006; Metzger, Thompson, \& Quataert 2006).
In all these scenarios, rapid rotation and the growth of strong magnetic 
fields due to compression, winding, and/or the MRI during the dynamical
phases of core collapse play essential roles.  Whether progenitors
boast the requisite rotation and initial seed fields remains to be determined
(Heger et al. 2004; Heger, Woosley, \& Spruit 2005), but an interesting subset
surely must.

To date, there has been a great deal of MHD code development in astrophysics,
mainly focused on simulating turbulent magnetic molecular 
clouds, accretion disks, AGN jets, solar and stellar dynamos,
and solar plasmas, but lately also on the question of gamma-ray bursts and supernovae. 
Stone \& Norman (1992ab) developed the ZEUS suite of codes that employs
the method of characteristics (MOC) to properly address MHD waves and
incorporates the constrained transport (CT) method (Evans \& Hawley 1988) to handle the divergence condition
on the B-field.  It includes gray radiation transport using a short-characteristics method.
Dai and Woodward (1994) developed an MHD Riemann solver and extended the piecewise parabolic method (PPM)
to multidimensional MHD equations. They also presented a PPM variant that exactly preserves
the conservation laws of magnetohydrodynamics together with the divergence-free condition (Dai \& Woodward 1998).
Gardiner \& Stone (2005) have followed ZEUS/MHD with a dimensionally-unsplit, 2nd-order, Godunov,
grid-based code (Athena) that combines the piecewise parabolic method (PPM) for spatial reconstruction
and the corner transport upwind (CTU) method of Colella (1990) for time-advancing 
multi-dimensional hydro with the CT method for divergence preservation.
Their numerical focus has been on making the CTU method compatible with 
both constrained transport and the finite-volume philosophy of PPM in the unsplit context.
The result is an accurate and conservative MHD code with great potential to address 
astrophysical problems. Smooth-particle hydrodynamics (SPH) codes have recently been augmented
to include magnetic fields, using divergence cleaning techniques (Price \& Monaghan 2005;
Ziegler, Dolag, \& Bartelmann 2006) and have been employed to study, among other
3D problems, the merger of binary neutron stars (Price \& Rosswog 2006).
Divergence cleaning techniques have been criticized vis \`a vis 
staggered-mesh, divergence-free formulations (Balsara \& Kim 2004) for artificially 
imposing post-facto cleaning methods to ensure $div B = 0$, but 
the approach is still useful for exploratory investigations. Aloy et al. 
(1999ab) have developed a sophisticated special-relativistic, axisymmetric
MHD code that incorporates as algorithmic elements finite-volume, 
Godunov, directional splitting, constrained transport, and an HLL-type 
solver.  Using a similar code, Obergaulinger et al. (2006) have recently 
investigated the MRI in the axisymmetric core-collapse context.

There are now several general-relativistic MHD codes that promise to
set the standard for astrophysical GRMHD simulations 
in the future.  Shibata \& Sekiguchi (2005) have constructed a GRMHD
code that features fully conservative numerical algorithms, dynamic space-times,
and the constrained transport method.  Noble et al. (2006) have built
a conservative GRMHD code employing approximate Riemann solvers and a
variant of constrained transport. Anninos, Fragile, \& Salmonson (2005)
have produced the GRMHD code, COSMOS++, that, while it also assumes a
fixed background space-time, has unstructured grids. COSMOS++ is non-Godunov,
uses artificial viscosity, uses finite-volume discretization, and has an AMR mode. 

Despite all these recent developments in computational MHD and enormous 
progress in computing power and in numerical techniques, there is still no single code
today that can simulate radiation-magnetohydrodynamics (RMHD) in full, mainly because a multi-angle,
multi-group transport code is expensive, even in 2D. In this paper, we
present a modest step toward such a code by reporting a new MHD scheme, incorporated 
into the already operational multi-group radiation hydrodynamics code VULCAN/2D (Livne 1993; 
Livne et al. 2004). Using multi-group, flux-limited diffusion we have obtained and recently published  
an interesting new explosion mechanism for core-collapse supernovae (Burrows et al. 2006ab).
We have also explored accretion-induced rotational collapse (Dessart et al. 2006a), 
convection in protoneutron stars (Dessart et al. 2006b), the angle-dependence of the 
neutrino field and neutrino heating on rotation rate (Walder et al. 2005), the mapping
between initial and final spins in the core-collapse context (Ott et al. 2006a), and
the gravitational radiation signatures of multi-D supernovae (Ott et al. 2006b).  
There are several basic advantages of VULCAN/2D over other codes (together with some drawbacks)
which enable the code to perform in difficult situations. One basic feature of VULCAN/2D is
 its ability to use unstructured grids, which in principle can efficiently cover any domain. 
 For example, polar grids become singular at the center of a star and Cartesian grids have difficulty
 resolving the center, while at the same time covering the entire star. In VULCAN/2D, we construct a grid
 by merging an inner curved rectangular with an outer polar grid. The resulting grid resolves the entire
 star with no singularity. In order to maintain this capability we have designed our MHD scheme for
 unstructured grids, having arbitrarily-shaped (but convex) quadrilateral cells.
 Another approach has been taken by Ardeljan et al. 2000, who
 use a 2D grid composed of triangles only, that also avoids the singularity of polar grids. 

In this paper, we report on the new MHD technique and present a number of test problems that 
demonstrate the ability of the code to simulate 2D magnetohydrodynamics problems
with both toroidal and poloidal fields. The paper is constructed as follows. In 
\S\ref{differ}, we describe in several subsections the basic equations and 
the finite-difference methodology for arbitrary geometry (\S\ref{basic}), 
the Lagrangean case (\S\ref{lagrange}), and the Eulerian case (\S\ref{euler}).
We continue in \S\ref{axial} with a discussion of the implementation in axially-symmetric 
geometries (\S\ref{axial}) and follow with a discussion of the method for evolving the 
toroidal component with rotation (\S\ref{toroid}).  We end \S\ref{differ} with a 
few words on necessary modifications to the Courant/CFL condition and boundary conditions (\S\ref{cfl}).
Section \ref{one} presents one-dimensional test problems and \S\ref{two} presents 
two-dimensional tests, some of them are new.  
For some of the test problems the magnetic stresses are comparable to the pressure stresses.
The associated tables and figures support our conclusion (\S\ref{conclusions})
that the automatically divergence-conserving MHD scheme we have implemented 
for general unstructured grids in our RMHD code VULCAN/2D provides competitive 
and reasonably accurate solutions to the dynamic equations of MHD in the Newtonian regime.

\section{The MHD Finite-Difference Scheme}
\label{differ}

In this paper we do not consider gravity, and therefore, omit gravity terms from the equations.
We also omit here all possible interactions between matter and radiation. This physics,
very relevant to the core-collapse problem, is already included in VULCAN/2D and has been reported
in several papers (Burrows et al. 2006ab; Ott et al. 2006ab, Dessart et al. 2006ab).
The MHD equations for incompressible flow in Lagrangean coordinates 
are (Landau \& Lifshitz 1960; Chandrasekhar 1961):
\begin{eqnarray}
 \Dert{\rho}+\rho\nabla\cdot\bv &=& 0   \\
 \rho\Dert{\bv} &=& -\nabla p +\frac{1}{4\pi}(\nabla \times \bB)\times \bB   \\
 \Dert{e} &=& -p\frac{d}{dt}(\frac{1}{\rho})                 \\
 \frac{d}{dt}\bBovr&=& \bBovr \cdot\nabla \bv \,.
\end{eqnarray}
Here, $e$ is the specific energy per gram and 
$p (\equiv p(\rho,e, Y_e)$) is the pressure, given by the equation of state
for a given density, $\rho$, internal specific energy per gram, $e$, and electron fraction, $Y_e$
(the mean number of electrons per baryon).
The Lagrangean time derivative $\frac{d}{dt}$ can be translated to the laboratory-frame by the transformation
 $\frac{d}{dt} = \frac{\partial}{\partial t}+\bv\cdot\nabla $.
 
The majority of current numerical schemes for hydrodynamics, including ours, use
cell-centered variables for the construction of finite-difference approximations to the fluid equations. 
But unlike Godunov-type schemes, like PPM, VULCAN/2D uses staggered differencing where the acceleration 
and velocity are computed at cell nodes rather than at cell centers. 
In such staggering schemes the acceleration is computed
by integrating the pressure times the normal over the surface of a control volume around each node.
 Therefore, we seek a scheme in which the magnetic field is also cell-centered, so that the contribution
of the magnetic stress to acceleration can be handled in a similar fashion.
This approach is different, for example, from the choices made in MOC-CT schemes (Stone \& Norman 1992b), where the 
magnetic field variables are face-centered (component $B_i$ is face-centered in the $i-$direction,
but cell-centered in the perpendicular direction). We do not calculate Lorentz forces directly but rather
compute the magnetic stress forces from the conservative form of the equations of motion. This
is achieved using standard integration schemes, as is done for example in Ardeljan (2000). 
One advantage of our approach is the exact conservation of momentum through the simulation.
In VULCAN/2D, we use the Cartesian coordinates  $(x,y)$ in the planar case and the
cylindrical coordinates  $(r,z)$  in the axisymmetric case. 
With these choices of coordinate systems, our scheme preserves linear momentum by construction, 
 ($p_x,p_y$ in the planar case, $p_z$ in the cylindrical case), and, in the latter case, 
angular momentum as well. We shall discuss the magnetic acceleration in more detail 
in section 2.2, where the cylindrical case is presented.  

\subsection{Differencing the Field Equations: Basic Approach for Arbitrary Geometry}
\label{basic}

Our field equation incorporates the {\it flux-freezing principle} used in Stone \& Norman (1992b).
For a given surface, $S$, enclosed by a closed loop, $C$, we define the flux:
\begin{equation} 
 \Phi = \int\int_{S} \bB \cdot \bn ds \,.
\end{equation}
If the surface moves with some arbitrary {\it grid velocity} $\bv_g$
the temporal derivative of $\Phi$ is given by
\begin{equation} 
 \der{\Phi}{t}= \oint_{C} (\bv-\bv_g)\times \bB dl \,.
\end{equation} 
Here, the time derivative is taken along the trajectory $\frac{d\mathbf x}{dt} = \bv_g$.
We can utilize eq.~(6) to evolve a cell-centered B-field in both Lagrangean
and Eulerian schemes in the following way. Take, for example, the two-dimensional
planar case, where a cell is some arbitrarily-shaped, convex, quadrilateral in the
$(x,y)$ plane. In order to evolve the field in time, according to the flux-freezing principle,
we need to define two control surfaces in that cell (the number of control surfaces must be equal to
the number of field components). There are several ways of constructing such control surfaces
inside a given cell. We considered two such ways -- first, connecting opposite mid-edge centers
and second, connecting diagonals of two opposite nodes. For each such choice we can construct a simple
scheme for the evolution of the magnetic field. However, the {\it diagonal scheme}
has been found to be more stable, while at the same time obeying the divergence-free condition of the magnetic field
in a very elegant way. Therefore, we shall focus on this choice in the rest of the paper.

\subsubsection{The Lagrangean Case}
\label{lagrange}

Ignoring for now the toroidal component of the field, $B_{\phi}$, the field is described by
its two poloidal cell-centered components $B_x,B_y$. By writing two finite-difference approximations
of eq.~(6) for the two surfaces one gets two linear equations for the two field components.
Let us consider first the Lagrangean case, where $\bv_g=\bv$ and the right-hand-side of eq. (6) vanishes.
Denote the two components of the two surfaces by $\Delta_{x,i}$ and $\Delta_{y,i}$, $i=1,2$.
Also, denote variables at the beginning of the timestep by superscript ${n}$ and variables at the
end of the timestep by superscript ${n+1}$. The discretization of eq. (6) in this case
yields the following two linear equations for the magnetic field at the end of the timestep:
\begin{eqnarray}
S_{x,1}^{n+1} B_x^{n+1} + S_{y,1}^{n+1} B_y^{n+1} &=& S_{x,1}^{n} B_x^{n} + S_{y,1}^{n} B_y^{n}  \\
S_{x,2}^{n+1} B_x^{n+1} + S_{y,2}^{n+1} B_y^{n+1} &=& S_{y,2}^{n} B_x^{n} + S_{y,2}^{n} B_y^{n} \,.
\end{eqnarray}
 The geometrical factors $S_{x,i},S_{y,i}$ are the components of the vector $\bn ds$.
\begin{eqnarray}
 S_{x,i} &=& -\Delta_{y,i} \\
 S_{y,i} &=& +\Delta_{x,i} \, .
\end{eqnarray}

It is easy to show that a divergence-free field remains divergence-free if we define
a finite difference {\it node-centered} approximation for the divergence. 
 From the divergence theorem, one gets the approximation 
\begin{equation} 
  div(\bB)(P) \approx \int\int_{S} \bB \cdot \bn ds / V
\end{equation}
where $P$ is a grid node, $S$ is a closed surface surrounding that point and $V$ is the volume contained
inside $S$. For our {\it diagonal scheme} the surface $S$ is composed of the four diagonals 
surrounding $P$ (see Fig.~1), and, therefore, the integral in eq. (11) is unchanged when the fields 
in the four relevant cells are evolved by eqs. (7)--(10).

\subsubsection{The Eulerian Case}
\label{euler}

In the more general case, where $\bv_g \ne \bv$, the right hand side of eq. (6) represents
advection of the B-field across the grid and does not vanish. We shall focus on the more 
common Eulerian case, where $\bv_g=0$. For each of our two surfaces we now have to compute the 
right hand side of eq. (6), and by this obtain two linear equations for the new cell-centered 
components of the field. To do this we need to extrapolate field values from cell centers 
to the four nodes of each cell. Note that for our {\it diagonal scheme} the divergence-free 
condition still holds throughout the evolution, because the advection terms cancel out in the integration 
of eq. (11) over a closed loop. This is true regardless of the {\it nodal}-field values. 
However, for reasons of stability and accuracy, these nodal values must be carefully defined according to 
monotonicity principles. This is a crucial step where the scheme is different from current MOC-CT 
schemes, which rely on one-dimensional flux-splitting in each direction. Our unstructured grid, 
which is certainly not orthogonal, requires more complex extrapolation in order to ensure
accuracy and stability.  

The Eulerian form of our difference scheme for the cell-centered components yields 
the following two linear equations:
\begin{eqnarray}
S_{x,1}^{n+1} B_x^{n+1} + S_{y,1}^{n+1} B_y^{n+1} &=& S_{x,1}^{n} B_x^{n} + S_{y,1}^{n} B_y^{n} 
           + \Delta t [B_x v_y - B_y v_x ]_1  \\
S_{x,2}^{n+1} B_x^{n+1} + S_{y,2}^{n+1} B_y^{n+1} &=& S_{y,2}^{n} B_x^{n} + S_{y,2}^{n} B_y^{n}
           + \Delta t [B_x v_y - B_y v_x ]_2 \,.
\end{eqnarray} 
Here, the brackets $[B_x v_y - B_y v_x ]_i$ stand for the difference between the enclosed 
expression on the two edges of diagonal $i$. The accuracy and stability of the scheme depend 
upon the exact definition of the nodal values of the magnetic field, while the velocity values 
are already node-centered. We, therefore, have to construct appropriate single-valued nodal field
components from the cell-centered basic field, and those will be used in the advection terms on the
right-hand-side of eqs. (12)--(13). The construction consists of the following steps:

\begin {itemize}
\item Compute spatial slopes for the field components according to van Leer's monotonic
   advection scheme (van Leer 1979).
\item Extrapolate cell-centered values to the nodes using those slopes. Note that
 for one-dimensional advection of a scalar field this procedure provides a monotonic and
 second-order-accurate scheme when the advection is done upwind. 
\item There are now four nodal values for each field component at any internal node $p$, and each
      of them is related to one diagonal surface in one of the surrounding cells (see diagonals a,b,c,d in Fig.~2).
      Using the velocity vectors at that node define the {\it upwind parameter} $\alpha_i=cos(\bv,\bf{T_i})$
      for each diagonal, where $\bf{T_i}$ is a unit vector along the  diagonal $i$ pointing to the
      node $p$. According to the sign of this {\it upwind parameter} define the type of
      each surrounding surface to be {\it donor-type} if $\alpha_i > 0 $ and {\it acceptor-type} otherwise.
\item Usually, there will be two {\it donor-type} surfaces (diagonals a and d in Fig.~2) and two {\it acceptor-type}
      surfaces (diagonals b and c in Fig.~2) at each (internal) node, but in regions where the grid is non-orthogonal
      this rule may not hold.
      Without loss of generality assume that surfaces $1,2$ are the two most {\it donor-type} surfaces out
      of the four surrounding surfaces, with nodal fields $\bB_1,\bB_2$ and normals $\bf N_1,\bf N_2$, respectively.
      Construct the single-valued nodal field vector $\bB_p$ by solving the two linear equations 
      $\bB_p \cdot \bf N_i = \bB_i \cdot \bf N_i \,$,  i=1,2,  which imply that the normal fluxes on the
      donor-type surfaces are the same when computed using $\bB_p$ or the primitive extrapolations $\bB_i$. 
\end {itemize}

 The above algorithm deserves a few remarks. First, the choice of donor-type surfaces actually guarantees
 that the two chosen diagonals (surfaces) are not parallel, so that this construction is indeed possible.
 Secondly,  the construction of the nodal magnetic vector from the donor-type normal components ensures that 
 the new fluxes are kept bound through  the advection step. This is essentially the heart of our generalization
 of van Leer's monotonic upwind scheme to the case of magnetic fluxes. Only minor changes are needed for boundary 
 points and we omit discussion here of those details. Finally, the same algorithm is applicable for general 
 nonzero {\it grid velocity}, which could be used in ALE simulations by VULCAN/2D.

\subsection{Implementation in Axially-Symmetric Geometry}
\label{axial}

Let us use the common coordinates, cylindrical $r$,$z$,$\phi$, for the radial, axial, and azimuthal 
directions. Assuming axial symmetry the azimuthal derivative of any variable vanishes 
everywhere, but the toroidal components of the magnetic field and the azimuthal velocity 
do not vanish. We first present the equations of motion in a form suitable for discretization:
\begin{eqnarray}
 \rho \Dert{v_r} &=& \frac{1}{4\pi}
       [B_z(\dpar{B_r}{z}-\dpar{B_z}{r})-\half\dpar{\Bf^2}{r}-\frac{\Bf^2}{r} ] - \dpdr \\
 \rho \Dert{v_z} &=& \frac{1}{4\pi}
       [\ovr\dpar{(rB_rB_z)}{r}+\half\dpar{(B_z^2-B_r^2-\Bf^2)}{z}] - \dpdz             \\
 \rho \Dert{(r\vf)} &=& \frac{1}{4\pi}[\ovr\dpar{(r^2B_r\Bf)}{r}+\dpar{(rB_z\Bf)}{z}] \,,
\end{eqnarray}
to be compared with Ardeljan et al. (2000) and chapter IX of Chandrasekhar (1961) for the incompressible inviscid
case. The main reason for writing the magnetic terms in this form is that the equation for the
axial momentum and the equation for the angular momentum have a conservative form. The difference
scheme of VULCAN/2D preserves these conservation laws by construction. The impulse and acceleration
at each grid point are obtained by integrating the above equations over the surface of a local
control volume, a cell in the staggered grid, which surrounds the point. Thus, momentum conservation laws are preserved
 exactly by the finite-difference scheme, as the integration provides at each surface segment two forces which have
 equal magnitude and opposite direction. Since this technique is standard in hydrodynamical schemes which use staggered
 grids we omit further details. Note however that the radial momentum in cylindrical coordinates is not a conserved
 quantity.

Returning to the evolution of the magnetic field we first discuss the poloidal components ($B_r, B_z$),
 while the evolution of the toroidal field will be described separately in \S\ref{toroid}.
The numerical approximation of those equations is obtained from the flux-freezing principle
in a way similar to the planar case. We first take a cell in the $(r,z)$ plane with the two
surfaces defined by the diagonals connecting opposite
nodes. The cell represents a torus around the axis of symmetry and the surfaces are actually
closed ring-like sheets centered around the axis of symmetry.
In the Lagrangean case, the resulting two linear equations for ($B_r,B_z$) are then

\begin{eqnarray}
S_{r,1}^{n+1} B_r^{n+1} + S_{z,1}^{n+1} B_y^{n+1} &=& S_{r,1}^{n} B_r^{n} + S_{z,1}^{n} B_y^{n}  \\
S_{r,2}^{n+1} B_r^{n+1} + S_{z,2}^{n+1} B_y^{n+1} &=& S_{r,2}^{n} B_r^{n} + S_{z,2}^{n} B_y^{n} \,,
\end{eqnarray} 
where the geometrical factors are 

\begin{eqnarray}
 S_{r,i} &=& -2 \pi r_i\Delta_{z,i} \\
 S_{z,i} &=& +2 \pi r_i\Delta_{r,i} \, ,
\end{eqnarray} 
$r_i$ being the radius at the center of each surface. We see that the only difference
with the planar case is the factor of $2\pi{r}$ multiplying each geometrical factor. Likewise, in 
the Eulerian case we get equations similar to eqs. (12)--(13) with appropriate $2\pi{r}$ factors. 
Therefore, the numerical technique for evolving the poloidal field is almost identical 
to that in the planar case.

\subsubsection{The Toroidal Component and Rotation}
\label{toroid}

For the time evolution of the toroidal component we need an extra equation. In the pure 
Eulerian case, which is our most common case, the equation for the toroidal field takes the form 
\begin{equation} 
\dpart{\Bf} = \frac{\partial}{\partial r}(B_r\vf-\Bf v_r)+\frac{\partial}{\partial z}(B_z\vf-\Bf v_z) \, .
\end{equation}
Using flux freezing again, we define an extra surface which extracts the toroidal field.
This is just the rectangular image of the cell in the $(r,z)$ plane. Integrating eq. (21) 
over this surface yields the desired equation. After integration, the right-hand-side 
gives edge-centered advection terms which we compute using the standard upwind van Leer scheme.

\subsubsection{Boundary Conditions and Time Steps for MHD Problems}
\label{cfl}

  We employ an {\it explicit} version of VULCAN/2D hydro and, thus, must restrict the
timestep according to stability constraints. 
We use the CFL condition (Courant \& Friedrichs 1948) updated to account for the propagation of
Alfv\'{e}n  and fast magnetosonic waves. In practice, we use
\begin{equation}
  \Delta t < \min \left( \frac{\Delta_x}{\mid v\mid + C_f} \right) \,,
\end{equation}
where $\mid$$v$$\mid$ is the local velocity (accounting for any rotational component as well)
and $C_f$ is the fast-magnetosonic speed ($C_f^2 = v_A^2 + C_s^2$), where $v_A$ is the Alfv\'{e}n
speed and $C_s$ the sound speed. Note that in core-collapse supernova
simulations, incorporating modest initial magnetic fields does not reduce noticeably the 
time increment, which after bounce is of the order of a microsecond. VULCAN/2D can also operate
in an implicit mode where the Courant condition for acoustic waves is removed. The MHD solver
however is explicit in nature. This implies that, depending on the field strength,
the timestep in future implicit simulations could be restricted by the speed of Alfv\'{e}n waves.

As to boundary conditions, they can be rather complicated if one wishes to simulate general MHD
problems in arbitrary geometries. Since we are interested mainly in the core-collapse problem,
and problems of a similar nature around compact objects, we choose to sacrifice the general case in 
favor of the most simple treatment. In our current implementation, the dynamical region in embedded
 in a large enough ambient region where either practically nothing changes during the simulations,
 or the outer boundary is not allowed to move. Therefore, under those restricted assumptions,
 a divergence-free initial field remain stationary in the outer regions of the flow. 

\section{Test Problems}
\label{tests}
 
A number of one-dimensional and two-dimensional test problems have been chosen to  
verify the code, in both planar and cylindrical geometries. 
We have computed some of the test problems suggested in Stone 
et al. (1992) and in Stone \& Norman (1992b). We have also designed a 
few new test problems which are relevant to dynamical collapse. 
The results of each VULCAN/2D simulation are compared to a simulation done under identical 
conditions using the publicly available MHD code ZEUS/2D (Stone et al. 1992; Stone \& Norman 1992b).

\subsection{One-Dimensional Test Problems}
\label{one}

 In this section, we present one-dimensional planar and cylindrical test problems.
For this purpose, we constructed a one-dimensional Lagrangean magnetohydrodynamic version,
VULCAN/1D, based upon the MHD formalism outlined in this paper and upon the earlier work
of Bisnovatyi-Kogan et al. (1976). VULCAN/1D is a variant that assumes that all the hydrodynamical
and magnetic variables are functions of ($r,t$) alone. We use this Lagrangean code for testing
our basic {\it flux freezing} approach in constructing the finite-difference scheme, and also for
post comparisons between VULCAN/2D and VULCAN/1D. In most cases, 
 the results of our VULCAN/2D (Eulerian) simulations are compared either to similar results
 obtained using  VULCAN/1D (Lagrangean) or to those obtained using ZEUS/2D (Eulerian).  

\subsubsection{Advection of a Magnetic Pulse}
\label{advection}

   This problem addresses the one-dimensional advection of a pulse of transverse magnetic field 
in Cartesian geometry, following the procedure presented in Evans \& Hawley (1988) and
in Stone et al. (1992). The velocity is fixed and uniform throughout the grid and the exact
solution is just a translation of the initial signal.
The test probes the magnitude of the numerical dispersion and diffusion of the 
advection scheme employed in the MHD solver, that should operate exactly like the same advection scheme
of a scalar field. 

   Our initial conditions are a one-dimensional grid of 500 zones, equally spaced
to tile the $x$-direction from 0 to 100. We use a Courant number of 0.5 and the equation 
of state is that of an ideal gas. Density and pressure are set to a fixed value 
everywhere, a value which does not matter in practice since
we perform pure advection., The velocity is unity everywhere, 
and we start with a non-zero transverse magnetic field 
between $x=10$ and $x=20$. Boundary conditions are periodic. 
We run the simulation until this pulse has advected five times its
initial width. We show in Fig.~\ref{fig_advec} the results for the ZEUS/2D simulations using the
donor-cell scheme and the van Leer scheme and for simulations using VULCAN/2D.
For comparison, we overplot the initial (square) pulse shifted along the x-direction by $v t$.
VULCAN/2D results are very close to ZEUS/2D with the van Leer scheme
(the two curves, essentially, overlapping). Given this good agreement, we do not investigate the 
distribution of the current density nor estimate errors as shown in Figs.~2--3 in Stone et al. (1992).
As expected, van Leer's scheme introduces a dramatic improvement of the results compared to donor cell.
Another higher-order variant, the PPA scheme of ZEUS/2D, improves only a little on van Leer's scheme,
while introducing non-monotonic unphysical oscillations (not shown here). 

\subsubsection {Brio-Wu Test Problem}
\label{briowu}

The Brio-Wu test problem (Brio \& Wu 1988) is a one-dimensional, planar, MHD Riemann problem.
Analytic solutions to MHD  Riemann problems are discussed in Ryu \& Jones (1995). The particular Brio-Wu problem 
belongs to a subclass of {\it compound states}, where the analytic solution is
ambiguous. We, therefore, compare our results to the numerical solution of ZEUS/2D. The $x$-range extends from 
0 to 100, and the initial parameters on the two sides of the initial discontinuity at $x=50$ are:

\begin{align*}   
\begin{cases}
x \le 50: &  \rho=1.000, P=1.0, B_{x}=0.75, B_{y}=+1  \\
x \ge 50: &  \rho=0.125, P=0.1, B_{x}=0.75, B_{y}=-1  \, .
\end{cases}
\end{align*}   

 The equation of state is of an ideal gas with $\gamma =2$. Boundary conditions are
 reflecting. To illustrate the behavior of VULCAN/2D {\it vis \`a vis} ZEUS/2D, we 
follow the same procedure found in Stone et al. (1992) and perform tests at different 
resolutions, assessing qualitatively and quantitatively the errors and how these vary with 
resolution. The Courant number for all simulations is set to 0.5.

 First, in Fig.~\ref{fig_bw_comp_var}, we present the solution of the Brio-Wu problem 
at $t=10$ computed with VULCAN/2D (black) and ZEUS/2D with the van Leer scheme (red).
The agreement seen in the graphs is excellent.
For a quantitative comparison of ZEUS/2D with the van Leer scheme, VULCAN/2D, and 
the analytic Brio-Wu results, we log in Table~\ref{tab_bw} the values predicted by each 
for the density, $\rho$, the pressure, $P$, the velocities, $V_x$ and $V_y$, and the 
B-field component, $B_y$. We provide these values at strategic locations in the 
fluid, which, from left to right, are the fast rarefaction (FR; $x=45$), the 
shock discontinuity (SC; $x=48$), the left and right sides of the contact discontinuity
(CD$_{\rm l}$ and CD$_{\rm r}$; $x=55$), the fast rarefaction (FR; $x=80$), as well
as the left and right initial boundary values for reference.

Following Stone et al. (1992), using self-convergence testing, we now estimate errors and 
convergence rates for both ZEUS/2D and VULCAN/2D in this specific Brio-Wu test.
The simulations presented in Fig.~\ref{fig_bw_comp_var} were done with $N = 512$ zones, uniformly 
spaced in the x-direction. We performed a series of similar calculations
with different resolutions: $N=$\,64, 128, 256, 1024, and 2048. 
We then calculated an $L_1$ error norm defined by 

\begin{equation}
L_1 = \left[ \frac{\int \mid   \rho_{2N}(x) -  \rho_{N}(x)\mid dx}{\int dx}   \right] \,, 
\end{equation}
where $\rho_{2N}(x)$ ($\rho_{N}(x)$) represents the density in the run with $2N$ ($N$) zones, 
both taken at $t=10$ and at the same location $x$. Note that here $L_1$ is a density error norm and 
that it has the units of a density, rather than being dimensionless. Using this measure, we seek the
rate at which the error varies with resolution, rather than the absolute level of that error.
Having six levels of resolution, we compute $L_1$ for five consecutive
pairs and show the results in the left panel of Fig.~\ref{fig_bw_error}. 
The convergence rate is similar for both simulations, i.e., ZEUS/2D with van Leer's scheme and VULCAN/2D. 

Finally, we display in the right panel of Fig.~\ref{fig_bw_error} the relative density
errors for the simulations at $N=256$ (broken) and $N=1024$ 
(solid) zones for VULCAN/2D (Black) and  ZEUS/2D/van Leer (red). 
Stone et al. (1992) noted that such an MHD shock-tube test, with a sharp
discontinuity, forces code accuracy to be first-order. Our 
comparisons between codes that are completely different in design provide 
an instructive look into code performance and the accuracy of the results generated.

\subsubsection{Propagation of an Alfv\'{e}n Wave in 1D}
\label{Alfven}

Here, we examine the propagation of a 1D Alfv\'{e}n wave, defined as a transverse perturbation on a 
given constant field $B_x=B_0$, embedded in a homogeneous medium of constant density $\rho$.
The solution of such a wave as a function of $x$ is naturally (Landau \& Lifshitz 1960):

\begin{equation}
v_{y}(x,t)=v_{0} cos(kx-\omega t)
\end{equation}

\begin{equation}
B_{y}(x,t)=-\sqrt{4\pi\rho}v_y \, .
\end{equation}

The perturbations $v_y$ and $B_y$ propagate along the $x$ direction with 
the Alfv\'{e}n speed $v_{A}=\frac{B_0}{\sqrt{4\pi\rho}}$.
We evolve the solution at $t=0$ with $k=2$ in the interval $[0,2\pi]$ for five periods and
compare the profiles at each integer period to the initial configuration. Table~\ref{tab_Alf} lists the
L1 error of $B_y$ for various resolutions ($N$) and 5 integer periods. 
As functions of resolution, our errors behave similarly to those 
presented in Table II of Toth (2000) for the traveling wave, and 
show a second-order rate of convergence. As a function of time, we observe 
a linearly growing error in each column in Table~\ref{tab_Alf} due to the 
fact that our scheme, like most other schemes, is only first-order accurate 
in time. Moreover, in more realistic problems which involve advection and/or acoustic
waves, second-order accuracy usually degrades to nearly first-order. 
This can be seen in Toth's Table II for the standing wave and in his Tables V-VIII, 
where the 7 tested schemes show a roughly first-order rate of convergence with spatial resolution. 

\subsubsection {Transport of Angular Momentum in a Rotating Disk }
\label{rotating}

This one-dimensional problem in cylindrical geometry tests the evolution of
a toroidal field, together with the transport of angular momentum in a differentially
rotating disk. Here, we start with differential rotation in the radial direction and
a corresponding radial magnetic field. This configuration is more typical of astrophysical problems
than is employed in the {\it Braking of aligned rotators} problem of  Stone et al. (1992). 
The stresses of the radial magnetic field transport angular momentum 
from the inner parts outward, evolving towards uniform
rotation, while a toroidal field grows from zero to significant magnitude.
Bisnovatyi-Kogan et al.(1976) suggested this mechanism for an MHD-driven supernova, but they need
extremely strong initial fields to generate an explosion.

We first set an initial disk configuration
in rotational equilibrium, where the centrifugal acceleration is balanced by pressure only, namely,
\begin{equation}
 \frac{1}{\rho}\dpdr = r\omega^{2}(r) \, .
\end{equation}
 Using an isentropic ideal gas with adiabatic index $\gamma$, $p=A\rho^{\gamma}$, and a rotation
law of the form 
\begin{equation}
 \omega(r) = \frac{\omega_{0} L^{2}}{L^{2} + r^{2}}  \, ,
\end{equation}
one can integrate the hydrostatic equation to get the entire disk configuration as a function of the radius.
We computed this test problem, with the parameters $p(r=0,t=0)=100, \rho(r=0,t=0)=1, \gamma=5/3, \omega_{0}=10$
 (in dimensionless units), using our one dimensional Lagrangean code VULCAN/1D and 
 VULCAN/2D in the Eulerian mode. The radius of the disk is $r_{max}=10$ and the width of
the initial rotation profile is $L=r_{max}/4$. In both simulations, we have used 500 radial zones and fixed, rigid,
outer boundary conditions. Without a magnetic field the disk remains static and stable in both
simulations, showing that the schemes do not introduce spurious numerical errors. For the real test we
turn on a radial magnetic field of magnitude $r B_{r} = 100$ and evolve the configuration until t=0.5.
Figure ~\ref{disk_plot} displays the toroidal field $B_\phi$ (left) and 
the angular velocity at three times (0.0,0.3,0.5).
The pluses, showing the results of VULCAN/2D, are plotted for 
only one fifth (100) of the zones in order not to hide
the solid lines. The excellent agreement between the two codes lends credence to the accuracy and
stability of our 2D Eulerian scheme, which is much more complicated than its 1D Lagrangean version.

\subsection{Two-Dimensional Test Problems}
\label{two}

There are only a few published dynamical test problems which are two-dimensional in nature.
Therefore, we constructed a number of such problems, which are close to astrophysical problems
in their dynamics, and compared the results of VULCAN/2D with those of ZEUS/2D. We focus on
problems which involve supersonic flow and strong shock waves, rather than on steady-state
problems like wind solutions. In the absence of analytic solutions code-to-code comparison is
the only way to validate multi-dimensional simulations. However, such comparisons should be done
with great care. In a few previous attempts, it was found that significant differences
between codes persist even when high-order schemes are used with maximum available resolution. 
Examples are the ``Santa Barbara Cluster Comparison Project'' (Frenk et al. 1999), the ``Richtmyer-Meshkov
Instability Study'' of Holmes et al. (1999) and the ``Rayleigh-Taylor Instability Study'' of the
``Alpha-Group'' (Dimonte et al. 2004). The differences between codes arise from the tendency of
unstable flows to form smaller and smaller structures due to vorticity generation. Therefore, one
can anticipate finding good agreement between two codes only on resolved structures, for which there are
at least 10-12 grid zones. In other words, there should be no point-wise convergence in simulations of
unstable flows in regions with small-scale fragmentation and shear (see also discussions by Calder \etal 2002).
Consequently, self-convergence tests in the sense described in \S\ref{briowu} are not a common practice in
multi-dimensional simulation. 

\subsubsection {Exploding Sphere/Blast Wave}
\label{explode}

This axisymmetric test problem consists of the expansion of a high pressure sphere
embedded inside an ambient cloud with initially uniform $B_z$ magnetic field. The specific
setup used is (in normalized dimensionless units): 

\begin{align*}
\begin{cases}
  0.01 < R < 0.2: & \rho=100, P=10^6, B_{r}=0, B_{z}=140  \\
  0.20 < R < 1.0: & \rho=1.0, P=0.60, B_{r}=0, B_{z}=140 \,,
\end{cases}
\end{align*}   
where $R$ is the spherical radius, and $B_r$ and $B_z$ refer to the cylindrical 
components of the magnetic field (spherical geometry is, however, used for the simulation).
The problem was evolved with an ideal gas equation of state with $\gamma=5/3$
and the Courant number used is again 0.5. The spherical grid is logarithmically spaced
in radius with $nr$ zones between 0.01 and 1.0 and uniformly spaced in angle with $nt$ zones between 
0 and $\pi$. Boundary conditions are reflecting and the simulation is stopped before the 
blast wave hits the outer boundary.

We show in Fig.~\ref{fig_exp} a montage of the density (top-left panel), pressure (top-right panel),
and the cylindrical $r$ and $z$ components of the magnetic field (bottom-left and bottom-right panels,
respectively) at the final time $t=0.0035$ in
a high-resolution simulation with $nr=280$ and $nt=240$. 
Each panel is broken into two halves, with the VULCAN/2D results on the right and the
ZEUS/2D results on the left. To give a better rendering of the dynamics and of the poloidal 
field morphology, we overplot velocity vectors (unsaturated; maximum length corresponds 
to a value of 165 or a physical length of 10\% of the width of the display). Magnetic field lines
are drawn starting from $z = -0.95\ r_{\rm max}$, equally spaced 
every 0.1$r_{\rm max}$ in the horizontal direction. The different paths 
followed by these lines between the two sides are indicative of 
differences in the magnetic field computed by VULCAN/2D and ZEUS/2D.
Overall, lines bunch up in the shock region where they accumulate due to flux freezing with the mass.

At the qualitative level, the agreement between the two codes is excellent. The 
colormaps of the components of the poloidal field show the compression of the 
vertical field $B_z$ and the generation of the radial component due to flux freezing.
Because the flow field of this problem is irrotational, we performed
 a resolution study in the sense described at the previous section. Since in this test problem there
 are only large-scale structures without shear we expect to see some kind of convergence.
 We show in Table~\ref{tab_exp} the results at the final time for the extrema of the density, 
pressure, $V_r$, $V_z$, $B_r$, and $B_z$. Figure~\ref{fig_exp} shows results for the 
highest-resolution simulation, while in Table~\ref{tab_exp} we give additional data for
low resolution ($nr=71$ and $nt=61$) and medium-resolution ($nr=141$ and $nt=121$) tests.
All quantities converge at higher resolution where they differ between the two codes
by no more than a few percent. However, at low resolution, differences can be of the order of
20-30\%. Note, however, that in general the low-resolution and 
medium-resolution values obtained by VULCAN/2D are closer to their high-resolution 
values than those obtained by ZEUS/2D. Using the highest resolution
 model as the reference model, the $L_1$ error norm 
for the density (see discussion in the Brio-Wu test above) gives a value for the low- and 
medium-resolution simulations, respectively, of 0.42 and 0.22 for ZEUS/2D, 0.24 and 0.19 for VULCAN/2D.   
Overall, ZEUS/2D and VULCAN/2D compare well for this explosion test, both qualitatively and 
quantitatively.

\subsubsection {Imploding Sphere}
\label{implode}

In this test, the initial spherically-symmetric configuration is split between  
an outer low-density, high-pressure shell and an inner high-density, low-pressure region.
The entire domain is threaded by a uniform vertical field $B_z$.
The initial setup is given by (in normalized dimensionless units):

\begin{align*}
\begin{cases}
  0.04 < R < 0.8: & \rho=10, P=1.00, B_{r}=0, B_{z}=40  \\
  0.80 < R < 2.0: & \rho=1.0, P=1000, B_{r}=0, B_{z}=40 \,.
\end{cases}
\end{align*}   

The problem was evolved with an ideal gas equation of state with $\gamma=5/3$,
and in the highest resolution version, we employ $nr=360$ logarithmically-spaced
radial zones and $nt=240$ uniformly-spaced angular zones on a full $\pi$ sector.
The initial configuration evolves so that the outer high-pressure region implodes,
compressing the inner region right down to the inner boundary at $R=0.04$, which then reexpands. 
The interaction of the magnetic field with the implosion leads to fragmentation of
the shock wave and to the formation of many small-scale structures.  
The ratio of the gas pressure to the magnetic pressure ($\beta$) reaches unity inside near bounce.
Note that for such tests with ZEUS/2D a large latitudinal flow developed 
at mid-latitude in the inner-boundary ``shell" promptly after the start 
of the simulation. We cured this anomaly in this otherwise static 
inner region by modifying the base boundary condition in ZEUS/2D
to enforce a null latitudinal velocity in the ghost zones and in the first active zone.

In a first test, we computed the implosion without rotation, for which we show
in Fig.~\ref{fig_imp_norot_t0pt05} results for the density (top left), pressure (top right),
$B_r$ (bottom left) and $B_z$ (bottom right) at time $t=0.05$
and in Fig.~\ref{fig_imp_norot_t0pt07} at time $t=0.07$.
The shock bounces at the inner rigid surface shortly after $t=0.05$, and we evolve the flow until $t=0.1$. 
Even before shock bounce, a complex pattern is formed due to the interaction between the
vertical magnetic field and the radially converging flow (Fig.~\ref{fig_imp_norot_t0pt05}). 
After bounce, the flow fragments on smaller scales and forms complex structures by non-radial
shock interactions (Fig.~\ref{fig_imp_norot_t0pt07}). The two codes
agree well on scales which are resolved by more than a few zones. Although VULCAN/2D is not designed
to conserve energy exactly (with or without magnetic fields) the total energy in this simulation is
conserved to an accuracy of $\sim$0.2 percent; the magnetic energy 
accounts for roughly 5 percent of the total.  
Both codes preserve accurately the left-right symmetry of $B_z$ 
and the anti-symmetry of $B_r$. Those symmetries
are seen in Fig.~\ref{fig_imp_norot_t0pt05} and subsequent 
figures, but can best be rendered by the morphology 
of the poloidal field lines (top right panels of Figs. \ref{fig_imp_norot_t0pt05} 
and \ref{fig_imp_norot_t0pt07} and in the right panels of 
Figs. \ref{fig_imp_rot} and \ref{fig_imp_highB}).

VULCAN/2D seems more dissipative than ZEUS/2D on the smallest scales and at sharp discontinuities. 
The slight differences at these smallest scales and at interfaces could be consequences of the fact 
that VULCAN/2D does not currently use any ``sharpening" techniques on discontinuities.
The agreement between the codes persists all the way to the end of the simulation at $t=0.1$ 
(not shown here). In Table~\ref{tab_impa}, we present the integrated energy components at several
times. We can see that the total energy is conserved at the same level of accuracy in the two codes.
The agreement between separate components of the energy budget is also good, where the largest
difference appears in the kinetic energy. Since this difference appears already at early times, and then
remains roughly constant, we speculate that the origin of this difference is a different treatment of
the initial shock wave that emerges from the large initial pressure discontinuity.

A second similar test uses the same initial input, but with initial rotation.
The initial rotation speed is given by $v_{\phi}(r) = \frac{\Omega_0 r}{1+(r/A)^2} $
and we choose $\Omega_0=10$ and $A=0.5$ ($r$ is the cylindrical radius).
In Fig.~\ref{fig_imp_rot}, we present comparisons of the angular velocity $\Omega$ (left column)
and the toroidal magnetic field $B_{\phi}$ (right column) before bounce ($t=0.05$; top row) 
and after bounce ($t=0.07$; bottom row). The initial field in this 
simulation is purely poloidal, so that the non-zero toroidal 
field stems from the winding of the poloidal component through differential
rotation. The complex 2D configuration of the angular velocity leads to both positive and negative
toroidal components of the magnetic field, depending on location (height above the inner boundary
and hemisphere). Again, the agreement between the codes is good for all structures
resolved by more than a few zones. In Table~\ref{tab_impb}, we present the integrated energy
 components at the times used for the non-rotating case. The conservation of total energy and the
agreement between different energy components are of the same quality as in the previous case. Note,
however, that while VULCAN/2D preserves the total angular momentum exactly ZEUS/2D does not, though
the deviation is only a few percent.

All simulations presented so far were set up with intermediate strength initial B-fields, and, indeed, comparable
magnetic and gas pressures were obtained only transiently in the simulations,  e.g., during the
bounce epoch in the implosion test. To test the high B-field case, we repeated the high-resolution non-rotating 
implosion test discussed earlier, but this time increased the initial magnetic field magnitude 
from 40 to 120 and 300. The magnetic field pressure then dominates the gas pressure. We present in
 Fig.~\ref{fig_imp_highB} a montage of 
colormaps of the density (left) and the quantity $\beta_{\rm plasma} = P/ (B^2/8\pi)$ (right), 
the ratio of thermal and magnetic pressures, for the cases with initial fields $B_z=120$ (top; $t=0.06$) and   
$B_z=300$ (bottom; $t=0.04$). Note how the material is more and more confined to motion along field lines
as the field strength increases, preventing any perpendicular motion in the highest $B_z$-value case.
Here, as before, differences are noticeable between VULCAN/2D and ZEUS/2D, for example in the extrema
reached by the density and $\beta_{\rm plasma}$ (or the gas pressure and the magnetic field components),
but overall the agreement is quite satisfactory. 
Simulations at high magnetic field strength imply large Alfv\'en speed and, thus, reduced timestep.
Because such simulations are costly, we stopped those simulations earlier than the final times of
the intermediate strength cases. 

\section{Conclusions}
\label{conclusions}

We have developed an ideal MHD scheme for general unstructured grids 
in two dimensions, plus rotation. The scheme consists of flux freezing 
of cell-centered magnetic vectors in both Lagrangean and Eulerian realizations.
For the Eulerian mode, we generalized the van Leer second-order advection scheme for cell-centered fluxes
in a unique way. The scheme is now incorporated into the (Newtonian) radiation hydrodynamics 
code VULCAN/2D and can be used for future two-dimensional RMHD simulations of 
rotating core-collapse supernovae and the early phases of gamma-ray bursts, as well as for 
simulations of a variety of other multi-D RMHD astrophysical problems. 
 
Via a number of one-dimensional and two-dimensional test problems, we find
that the accuracy of the scheme is comparable to that of high-order MOC-CT schemes
used, for example, in the code ZEUS/2D. It is, however, impossible to prove a 
{\it formal order of accuracy} for our scheme because it does not consist of one-dimensional sweeps. 
In actual test problems we see that VULCAN/2D is somewhat more dissipative on small, 
unresolved, scales than ZEUS/2D, but the agreement between the two codes on structures that are 
resolved by more than a few zones is good. The extra dissipation of VULCAN/2D is 
probably related to numerical details that are not related to the MHD scheme, 
and do not affect resolved scales. The MHD scheme is designed to conserve the nodal
finite-difference representation of  $div(\bB)$ automatically to machine accuracy at all times.
Moreover, it conserves linear and angular momenta in the axisymmetric cylindrical case. 

\acknowledgments

We thank Mordecai Mac Low, Christian Ott, and Jeremiah  Murphy for interesting discussions.
This research was supported by The Israel Science Foundation (grant No. 805/04).
We also acknowledge support from the Scientific Discovery through Advanced Computing
(SciDAC) program of the DOE, under grants DE-FC02-01ER41184 and DE-FC02-06ER41452,
and from the NSF under grant AST-0504947. This research used resources of the National
Energy Research Scientific Computing Center, which is supported by the
Office of Science of the U.S. Department of Energy under Contract No. DE-AC03-76SF00098.
Finally, this research used resources of the NCCS at Oak Ridge National Laboratory,
which is supported by the Office of Science of the Department of 
Energy under contract DE-AC05-00OR22725.

\newpage

\begin{deluxetable*}{ccccccccc}
\tablewidth{15cm}
\tabletypesize{\scriptsize}
\tablecaption{Values of selected quantities for the Brio-Wu test at $t=10$.
\label{tab_bw}}
\tablehead{
\colhead{Variable}&
\colhead{Source}&
\colhead{Left}&
\colhead{FR}&
\colhead{SC}&
\colhead{CD$_{\rm l}$}&
\colhead{CD$_{\rm r}$}&
\colhead{FR}&
\colhead{Right}
}
\startdata
$\rho$ & VULCAN/2D  &  1.0    & 0.664 & 0.839  &  0.704  & 0.242  & 0.116  &   0.125  \\
       & ZEUS/2D    &  1.0    & 0.660 & 0.840  &  0.699  & 0.250  & 0.116  &   0.125  \\
       & BW         &         & 0.676 &        &  0.697  &        &        &          \\
\\
\hline
\\
$P$    & VULCAN/2D  &  1.0    & 0.441 & 0.719  &  0.505  & 0.505  & 0.086  &   0.1  \\
       & ZEUS/2D    &  1.0    & 0.435 & 0.712  &  0.501  & 0.501  & 0.086  &   0.1  \\
       & BW         &         & 0.457 &        &  0.516  &        &        &          \\
\\
\hline
\\
$V_x$  & VULCAN/2D  &  0.0    &  0.663  & 0.485  &  0.596  & 0.596  & -0.264  &   0.0  \\
       & ZEUS/2D    &  0.0    &  0.672  & 0.488  &  0.603  & 0.603  & -0.279  &   0.0  \\
       & BW         &       &  0.637  &        &  0.599  &        &         &       \\
\\
\hline
\\
$V_y$  & VULCAN/2D  &  0.0    & -0.247   & -1.19  & -1.58  & -1.58   &  -0.185 &   0.0  \\
       & ZEUS/2D    &  0.0    & -0.251   & -1.16  & -1.58  & -1.58   &  -0.196 &   0.0  \\
       & BW         &       & -0.233   &        & -1.58  &         &         &       \\
\\
\hline
\\
$B_y$  & VULCAN/2D  &  1.0  & 0.568   & -0.190  & -0.537  & -0.537  & -0.892  &   -1.0  \\
       & ZEUS/2D    &  1.0  & 0.562   & -0.195  & -0.539  & -0.539  & -0.886  &   -1.0  \\
       & BW         &       & 0.585   &         & -0.534  &         &         &       \\
\enddata
\tablecomments{Hydromagnetic variables at selected locations in the BW solution, which, from left to right,
 are the fast rarefaction (FR; $x=45$), the shock discontinuity (SC; $x=48$),
 the left and right sides of the contact discontinuity (CD$_{\rm l}$ and CD$_{\rm r}$; $x=55$),
 the fast rarefaction (FR; $x=80$), as well as the left and right initial boundary values for reference
 (see Fig.\ref{fig_bw_comp_var}). Note how these values closely match.
}
\end{deluxetable*}

\newpage
\begin{deluxetable*}{ccccc}
\tablewidth{15cm}
\tabletypesize{\scriptsize}
\tablecaption{L1 errors in $B_y$ for the Alfv\'{e}n Wave test
\label{tab_Alf}}
\tablehead{
\colhead{Periods}&
\colhead{N=16}&
\colhead{N=32}&
\colhead{N=64}&
\colhead{N=128}
}
\startdata
 P=1  &  0.0086 & 0.0021  & 0.00054  & 0.00015  \\
\\
 P=2  &  0.0171 & 0.0041  & 0.00106  & 0.00028  \\
\\
 P=3  &  0.0256 & 0.0062  &  0.00158 & 0.00041  \\
\\
 P=4  &  0.0339 & 0.0083  &  0.00210 & 0.00054  \\
\\
 P=5  &  0.0423 & 0.0103  & 0.00262  & 0.00068  \\
\enddata
\tablecomments{The value of N indicates the number of zones per wavelength}

\end{deluxetable*}

\newpage
\begin{deluxetable*}{cccccccc}
\tablewidth{15cm}
\tabletypesize{\scriptsize}
\tablecaption{Values of selected quantities for the explosion test at $t=0.0035$, using low, medium,
and high resolution.
\label{tab_exp}}
\tablehead{
\colhead{Variable}&
\colhead{Code}&
\multicolumn{2}{c}{Low resolution}&
\multicolumn{2}{c}{Medium resolution}&
\multicolumn{2}{c}{High resolution}\\
\hline
\\
\colhead{}&
\colhead{}&
\colhead{Min.}&
\colhead{Max.}&
\colhead{Min.}&
\colhead{Max.}&
\colhead{Min.}&
\colhead{Max.}
}
\startdata
$\rho$ & VULCAN/2D  & 1.00   &  3.53 & 1.0    &   4.48  &  1.0  &  5.4 \\
       & ZEUS/2D    & 1.02   &  2.83 & 0.99   &   4.12  &  0.97 & 5.3  \\
\\
\hline
\\
$P$    & VULCAN/2D  & 0.60   &10157  & 0.6    &  10119  &  0.6  &  10525 \\
       & ZEUS/2D    & 2.18   & 7995  & 0.59   &  10186  &  0.57 &  10857 \\
\\
\hline
\\
$V_r$  & VULCAN/2D  &  -0.18 &  152.9 & -0.15  &  157.6  & -0.1  &  161.2 \\
       & ZEUS/2D    &  0.0  &  143.5 & 0.26   & 156.6   & 0.18  &  160.9 \\
\\
\hline
\\
$V_z$  & VULCAN/2D  & -161.2 &  161.2 & -164.0 &  164.0  & -164.5  &  164.5  \\
       & ZEUS/2D    & -146.3 &  146.3 & -158.8 &  158.8  & -164.3   & 164.3  \\
\\
\hline
\\
$B_r$  & VULCAN/2D  & -160.7 &  160.7 & -205.8 & 205.8   & -214.0  &  214.0 \\
       & ZEUS/2D    & -127.7 &  127.7 & -189.3 & 189.3   & -208.0  &  208.0 \\
\\
\hline
\\
$B_z$  & VULCAN/2D  & -0.8   &  381.3 & -3.18  &   405.9  & -4.5  &  417.0 \\
       & ZEUS/2D    & -7.8   &  331.1 & -6.3   &   402.2  & -8.6  &  420.0  \\
\enddata

\end{deluxetable*}

\newpage

\begin{deluxetable*}{cccccccc}
\tablewidth{15cm}
\tabletypesize{\scriptsize}
\tablecaption{Energy components in the high-resolution non-rotating implosion test.
\label{tab_impa}}
\tablehead{
\colhead{Time}&
\colhead{Code}&
\colhead{Ekin}&
\colhead{Eint}&
\colhead{Emag}&
\colhead{Erot}&
\colhead{J   }&
\colhead{Etot}
}
\startdata
\\
  0. &  VULCAN  &  0.0  &   \E{4.705}{4}  &  \E{2.133}{3}  & 0. & 0.  &  \E{4.918}{4} \\
     &  ZEUS    &  0.0  &   \E{4.705}{4}  &  \E{2.117}{3}  & 0. & 0.  &  \E{4.916}{4} \\
\hline
\\
  0.03  &  VULCAN  &  \E{6.575}{2}  &  \E{4.625}{4} &  \E{2.223}{3} & 0. & 0.  & \E{4.913}{4} \\
        &  ZEUS    &  \E{6.954}{2}  &  \E{4.613}{4} &  \E{2.216}{3} & 0. & 0.  & \E{4.905}{4} \\
\hline
\\
  0.05  &  VULCAN  &  \E{6.598}{2}  &  \E{4.618}{4} &  \E{2.270}{3}  & 0. & 0.  & \E{4.911}{4} \\
        &  ZEUS    &  \E{6.961}{2}  &  \E{4.605}{4} &  \E{2.267}{3}  & 0. & 0.  & \E{4.902}{4} \\
\hline
\\
  0.07  &  VULCAN  &  \E{1.598}{2}  & \E{4.658}{4}  &  \E{2.353}{3}  & 0. & 0.  & \E{4.910}{4} \\
        &  ZEUS    &  \E{1.881}{2}  & \E{4.645}{4}  &  \E{2.364}{3}  & 0. & 0.  & \E{4.900}{4} \\
\hline
\\
  0.10  &  VULCAN  &  \E{6.247}{2}  & \E{4.624}{4}  &  \E{2.215}{3}  & 0. & 0.  & \E{4.908}{4} \\
        &  ZEUS    &  \E{6.469}{2}  & \E{4.611}{4}  &  \E{2.216}{3}  & 0. & 0.  & \E{4.898}{4} \\
\enddata

\end{deluxetable*}

\newpage

\begin{deluxetable*}{cccccccc}
\tablewidth{15cm}
\tabletypesize{\scriptsize}
\tablecaption{Energy components in the high-resolution rotating implosion test.
\label{tab_impb}}
\tablehead{
\colhead{Time}&
\colhead{Code}&
\colhead{Ekin}&
\colhead{Eint}&
\colhead{Emag}&
\colhead{Erot}&
\colhead{J   }&
\colhead{Etot}
}
\startdata
\\
  0. &  VULCAN  &  0.0  &  \E{4.705}{4}  &  \E{2.133}{3}  & \E{1.076}{2} &  87.08  &  \E{4.929}{4} \\
     &  ZEUS    &  0.0  &  \E{4.705}{4}  &  \E{2.117}{3}  & \E{1.076}{2} &  87.08  &  \E{4.927}{4} \\
\hline
\\
  0.03  &  VULCAN  &  \E{6.199}{2}  &  \E{4.626}{2}  &  \E{2.222}{3}  & \E{1.357}{2} & 87.08  & \E{4.923}{4}  \\
        &  ZEUS    &  \E{6.546}{2}  &  \E{4.615}{2}  &  \E{2.214}{3}  & \E{1.362}{2} & 87.70  & \E{4.916}{4}  \\
\hline
\\
  0.05  &  VULCAN  &  \E{5.861}{2}  &  \E{4.617}{4}  &  \E{2.255}{3}  & \E{1.988}{2} & 87.08  & \E{4.921}{4}  \\
        &  ZEUS    &  \E{6.172}{2}  &  \E{4.606}{4}  &  \E{2.253}{3}  & \E{2.021}{2} & 88.68  & \E{4.913}{4}  \\
\hline
\\
  0.07  &  VULCAN  &  \E{1.693}{2}  &  \E{4.645}{4}  &  \E{2.323}{3}  & \E{2.610}{2} & 87.08  & \E{4.920}{4}  \\
        &  ZEUS    &  \E{1.869}{2}  &  \E{4.632}{4}  &  \E{2.332}{3}  & \E{2.733}{2} & 90.16  & \E{4.912}{4}  \\
\hline
\\
  0.10  &  VULCAN  &  \E{5.811}{2}  & \E{4.623}{4}   &  \E{2.217}{3}  & \E{1.600}{2} & 87.08  & \E{4.919}{4}  \\
        &  ZEUS    &  \E{6.037}{2}  & \E{4.611}{4}   &  \E{2.213}{3}  & \E{1.737}{2} & 93.10  & \E{4.910}{4}  \\
\enddata

\end{deluxetable*}

\clearpage
\newpage

\begin{figure}
\epsfig{file=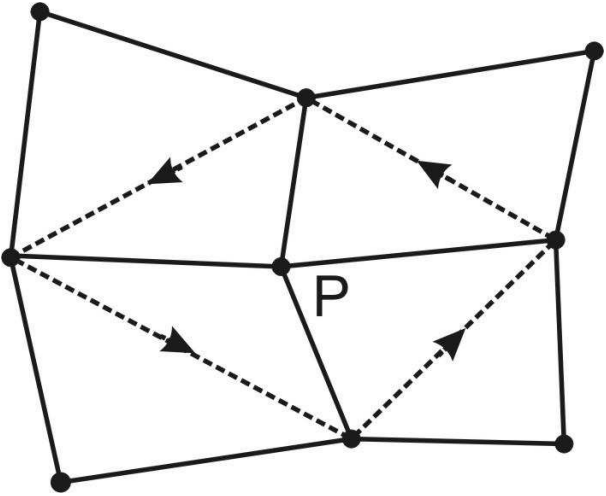,width=8cm}
\caption{
Scheme template: solid line quadrilaterals are grid cells.
The dashed-line closed loop around P is the control surface for computing $div(\bB)$ at P.
In each cell there are two diagonals used to compute $(B_r,B_z)$.
}
\end{figure}


\begin{figure}
\epsfig{file=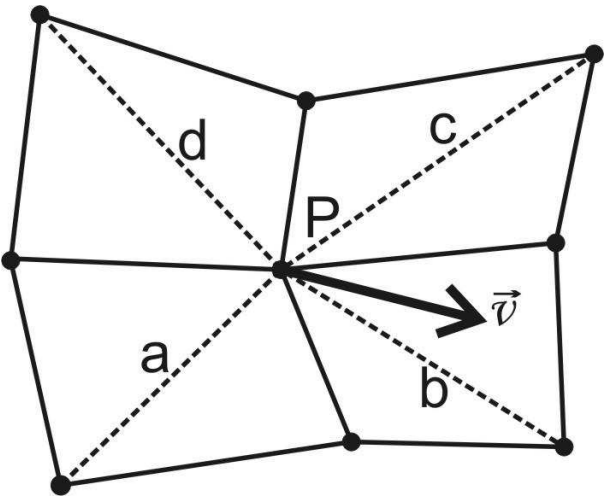,width=8cm}
\caption{
Template of the advection step: the velocity vector defines
two {\it donor-type} diagonals (a,d) and two {\it acceptor-type} diagonals (b,c),
according to the upwind stability criterion.}
\end{figure}

\clearpage

\begin{figure}[tp!]
\epsfig{file=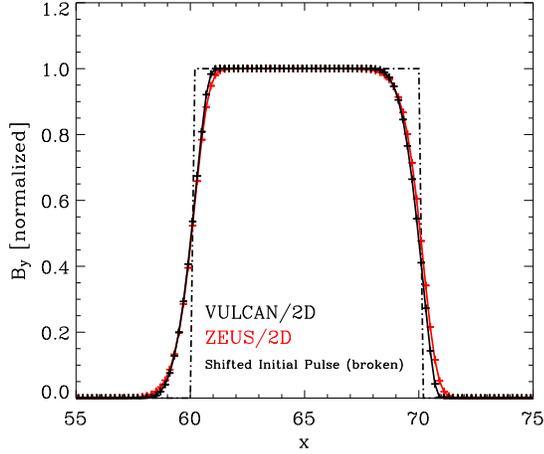,width=8cm}
\caption{Illustration of the advected $B_y$--field for the pure advection test of a transverse
magnetic pulse at $t=50$, employing 500 zones, using the ZEUS code (green: Donor cell scheme;
red: van Leer scheme) and VULCAN/2D (solid black). For comparison, we overplot
the ``exact'' advection (translation) of the initial pulse (broken black).}
\label{fig_advec}
\end{figure}


\begin{figure}[tp!]
\epsfig{file=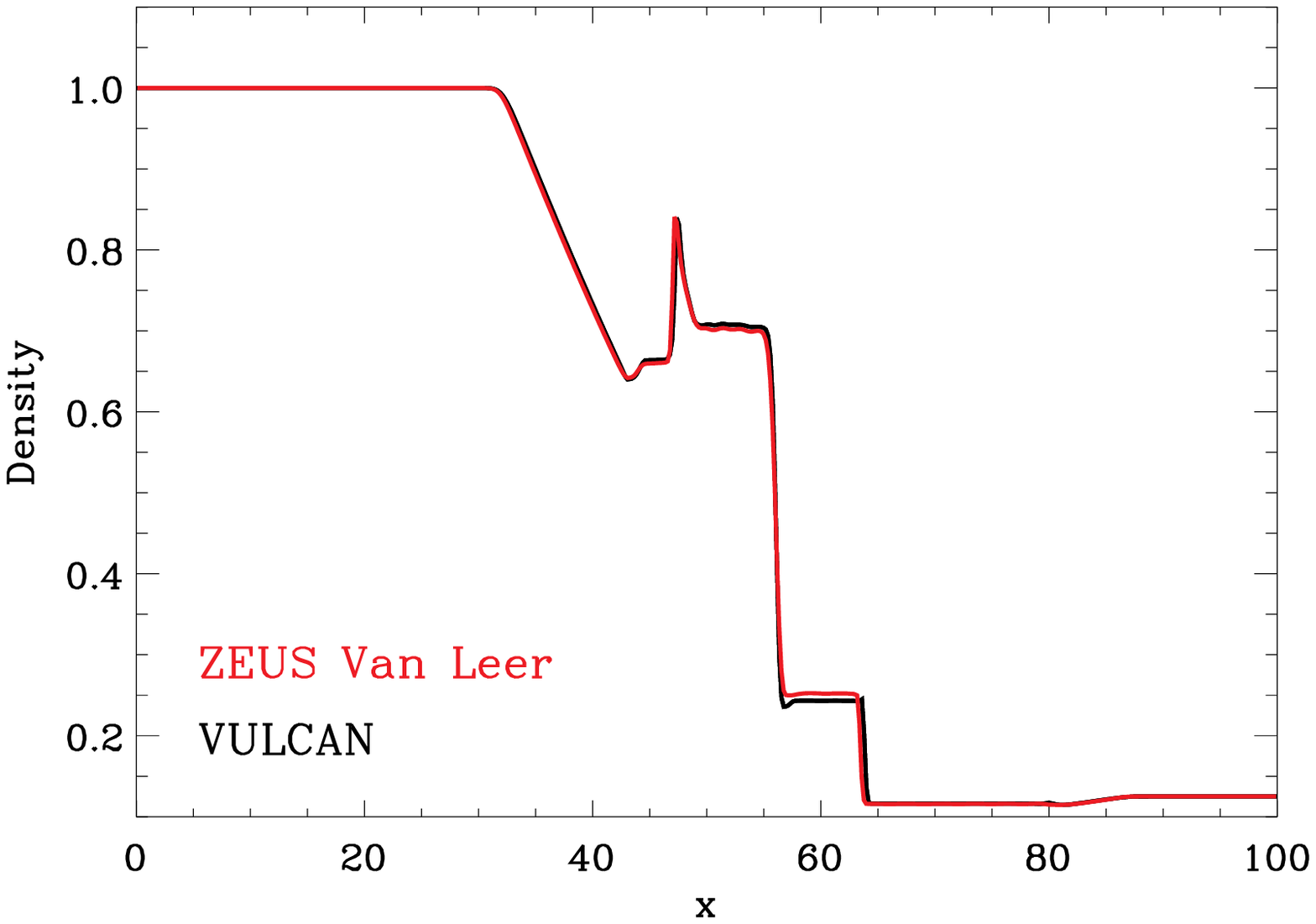,width=6cm}
\epsfig{file=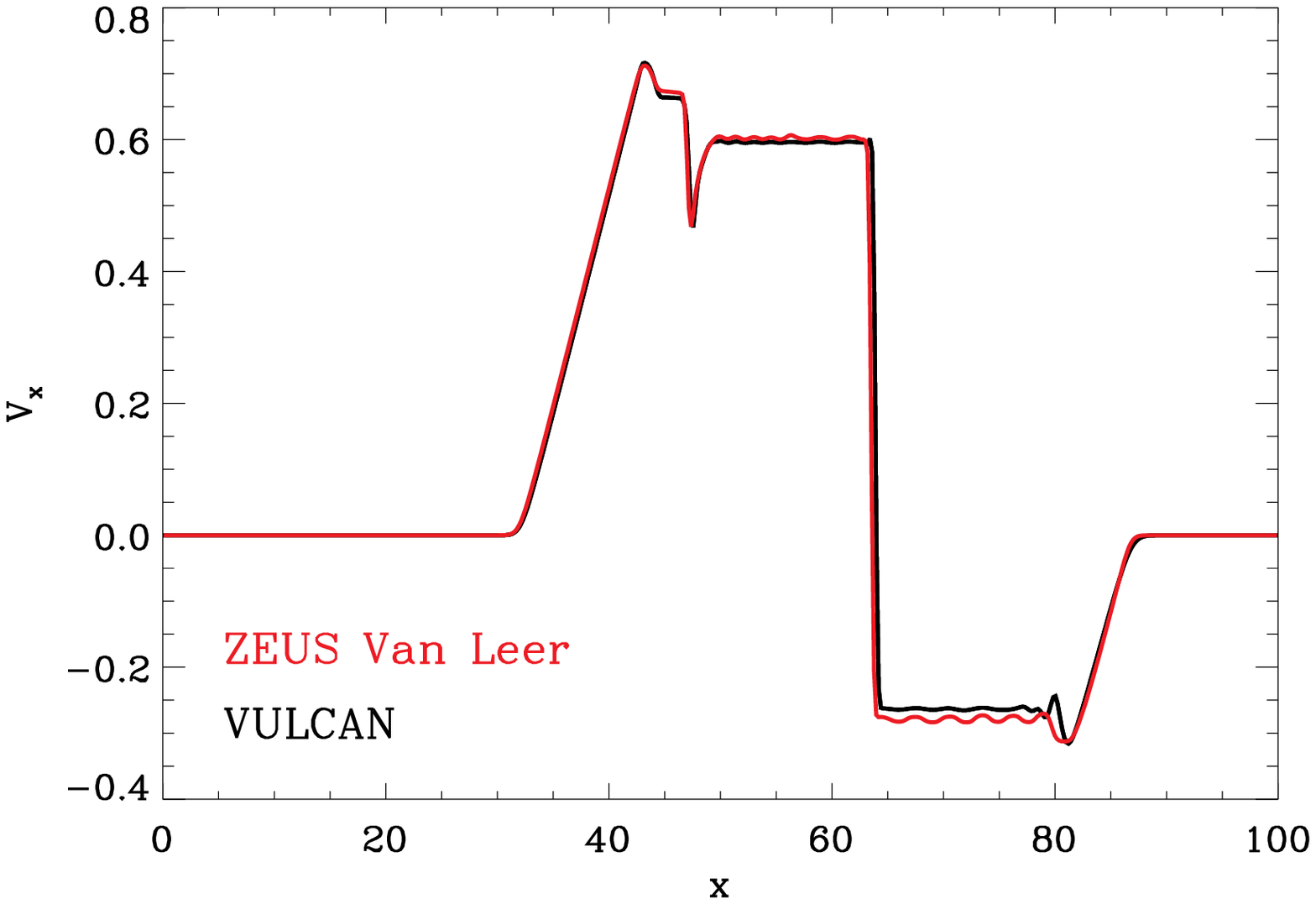,width=6cm}
\epsfig{file=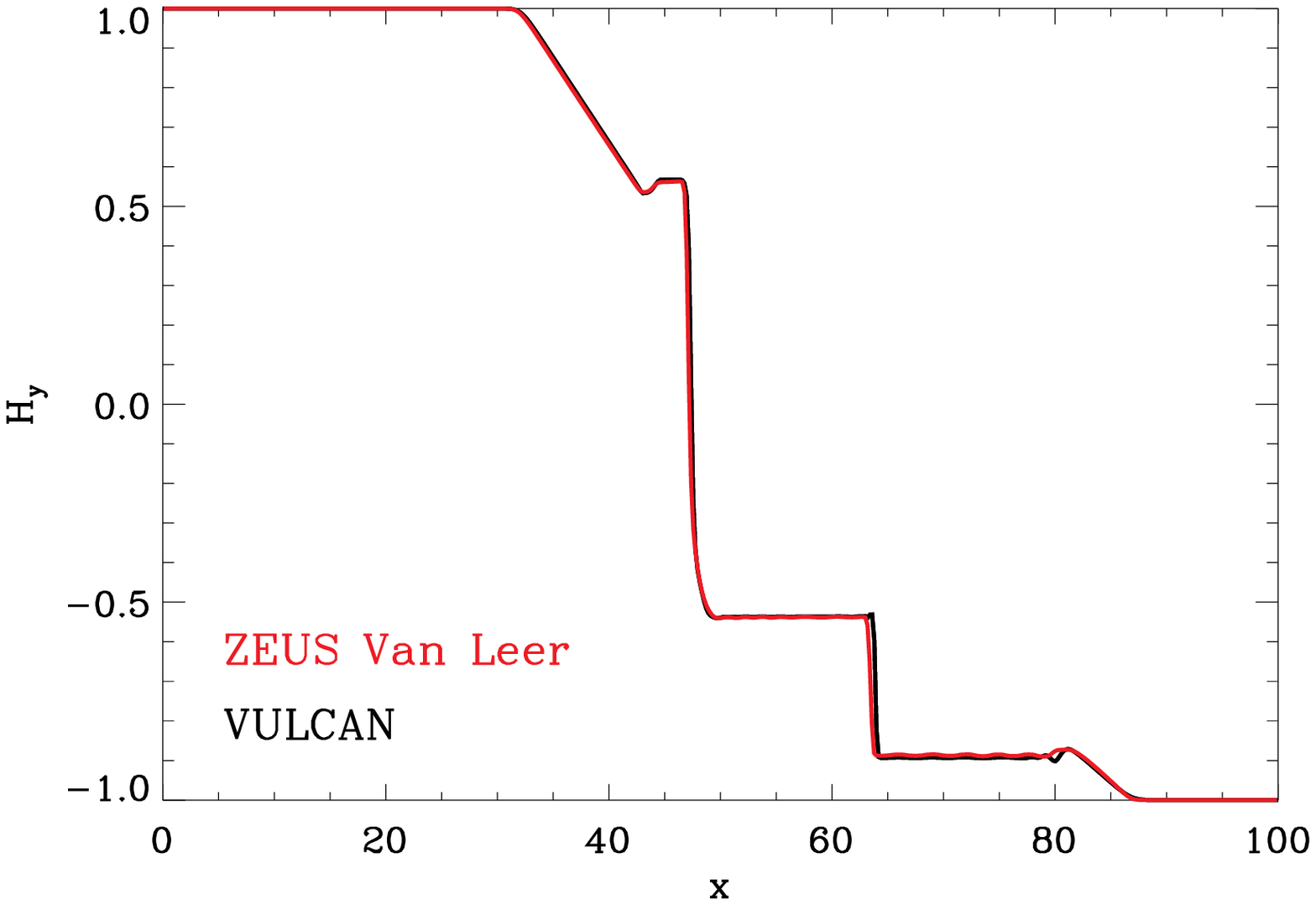,width=6cm}
\caption{Illustration of the density (top), x-velocity (middle), and $B_y$-field (bottom)
for the Brio-Wu test at $t=10$, employing 512 zones, using VULCAN/2D (black) and ZEUS/2D
(red).}
\label{fig_bw_comp_var}
\end{figure}


\begin{figure}
\plottwo{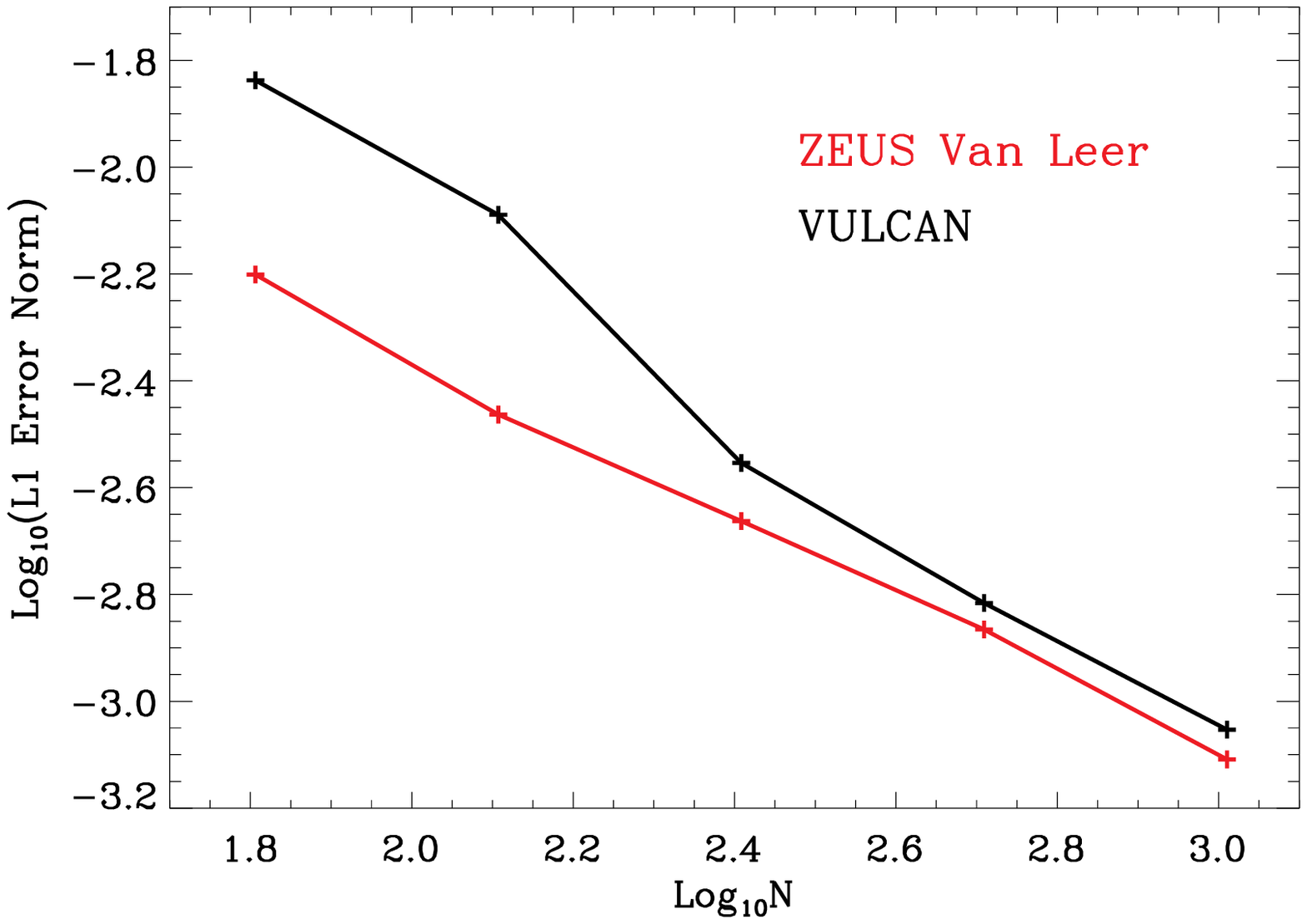}{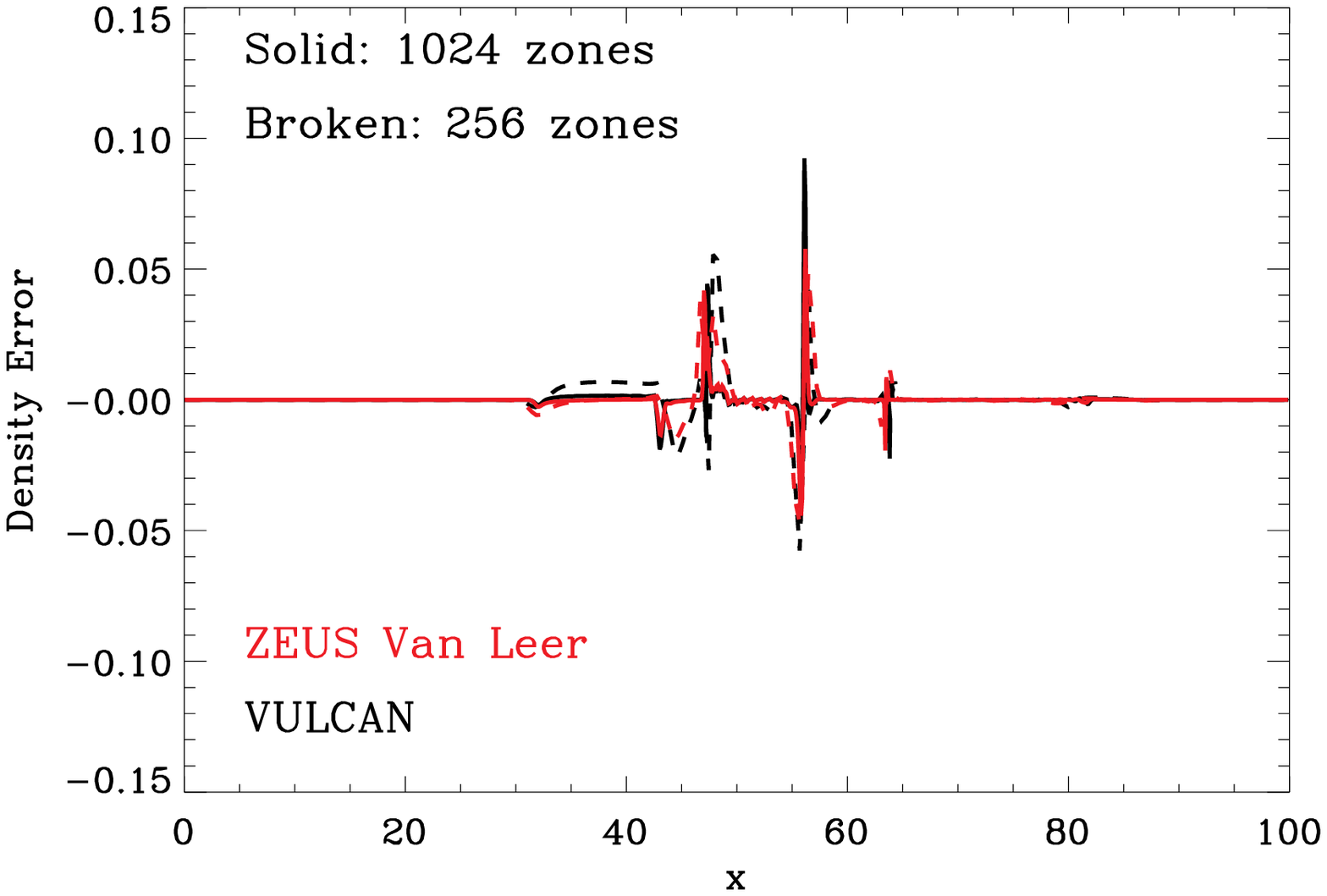}
\caption{{\it Left:} Illustration of the $L_1$ error norm computed on the density for the
Brio-Wu test with VULCAN (black) and ZEUS/2D (red) as a function of the
logarithm of the number of zones employed, {\it i.e.}  $N=$64, 128, 256, 512, 1024, and 2048.}
\label{fig_bw_error}
\end{figure}

\begin{figure}
\plottwo{f6a.eps}{f6b.eps}
\caption{ \hspace{1.5cm} {\it Left:} The toroidal field 
$B_\phi$; \hspace{4cm} {\it Right}: The angular velocity
\newline 
 Solid lines - VULCAN/1D Lagrangean simulation, pluses - VULCAN/2D Eulerian simulation}
\label{disk_plot}
\end{figure}

\clearpage

\begin{figure*}[tp!]
\epsscale{1.1}
\plottwo{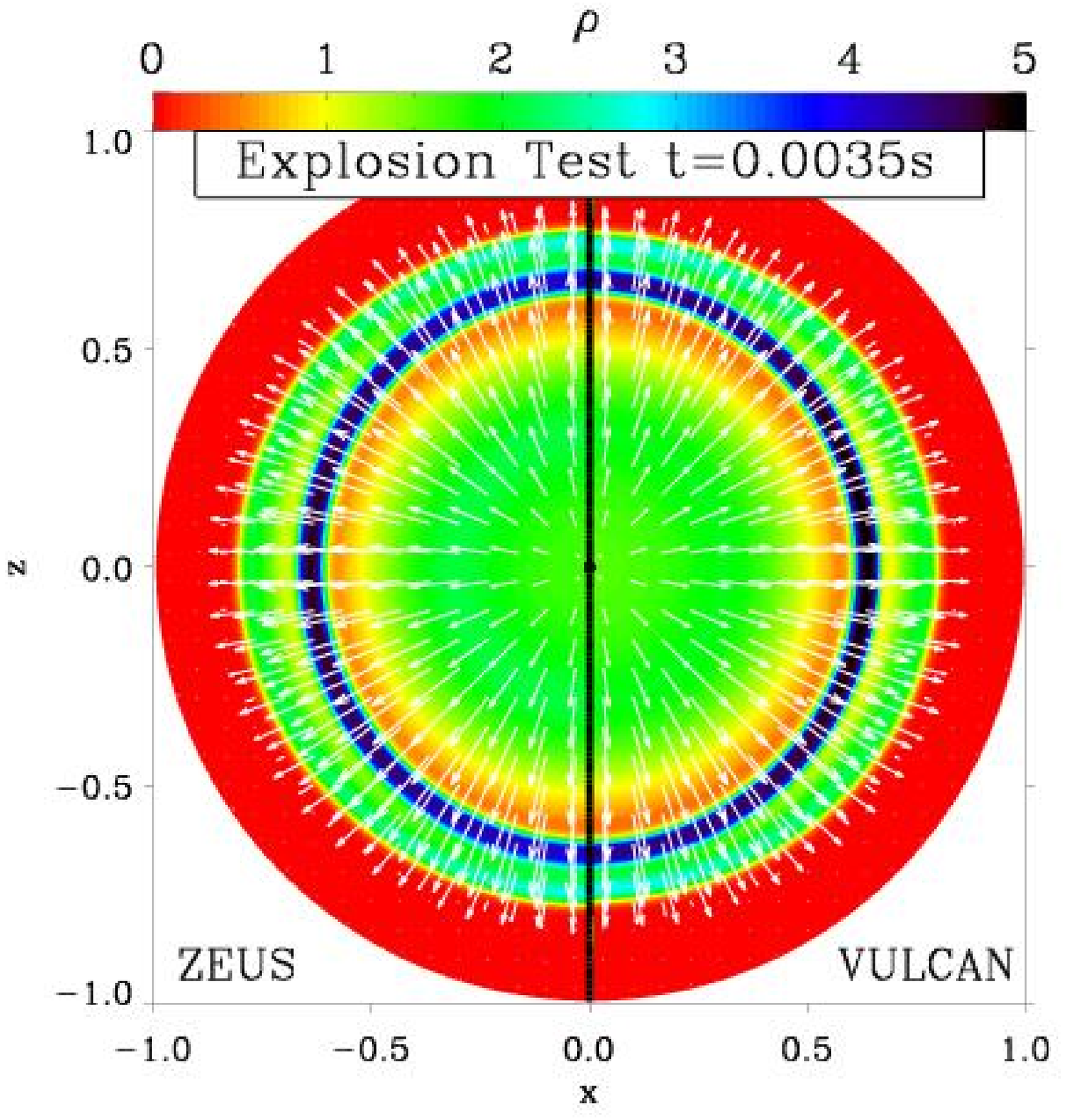}{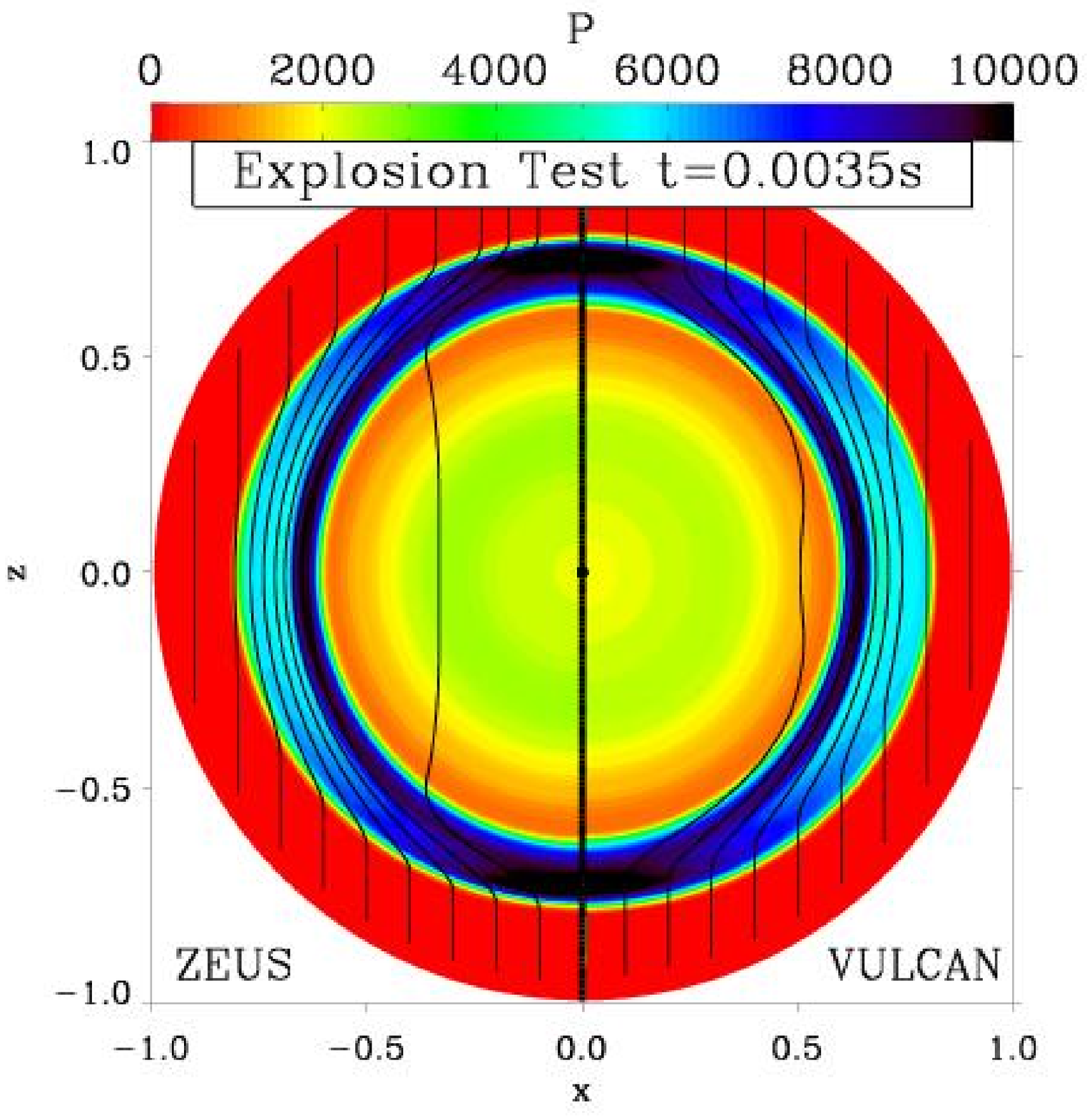}
\plottwo{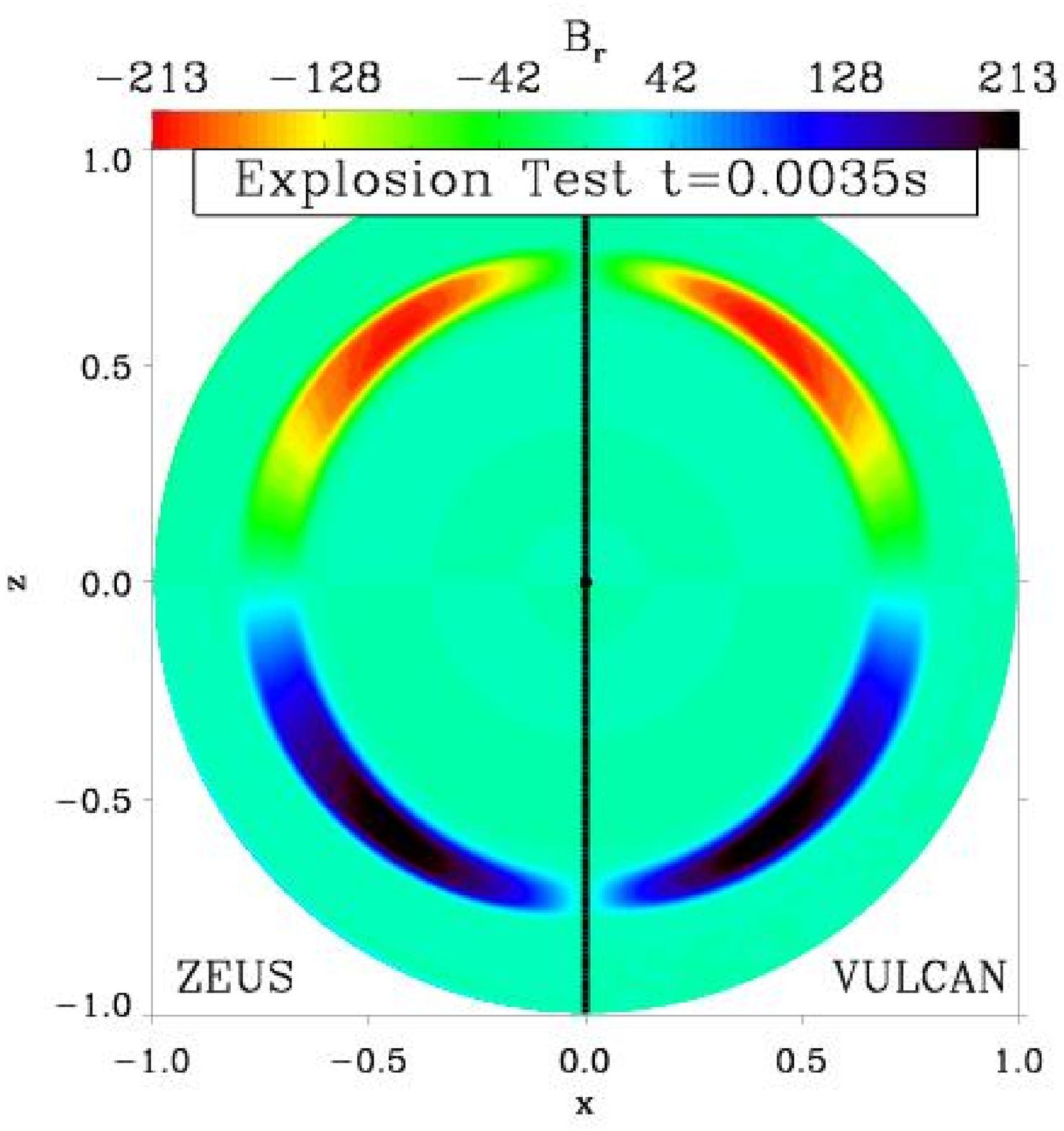}{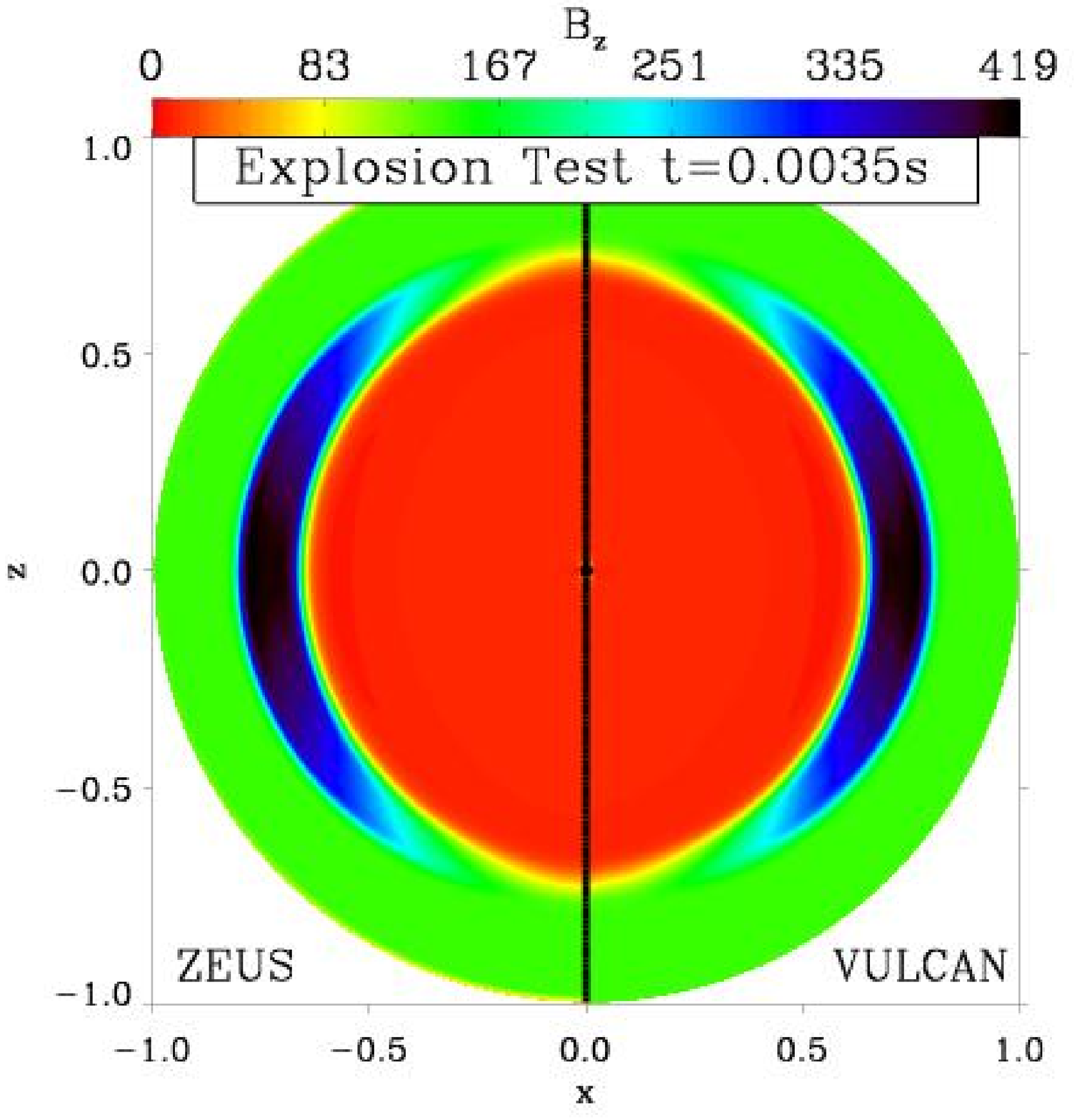}
\caption{Colormaps of the high-resolution explosion test at t=0.035: density (top left),
 pressure (top right), $B_r$ (bottom left) and $B_z$ (bottom right).
 Each panel shows the ZEUS/2D (left) and VULCAN/2D (right) results.
 Velocity vectors are overplotted in white in the top-left box, with a maximum length
 set to 10\% of the width of the panel, and a maximum magnitude of 165 (all quantities are dimensionless).
 Magnetic field lines are overplotted in black in the top-right box,
 with footpoints at $z = -0.95 r_{\rm max}$ equally spaced along $x$ every 0.1$r_{\rm max}$.
}
\label{fig_exp}
\end{figure*}

\clearpage

\begin{figure*}[tp!]
\epsscale{1.1}
\plottwo{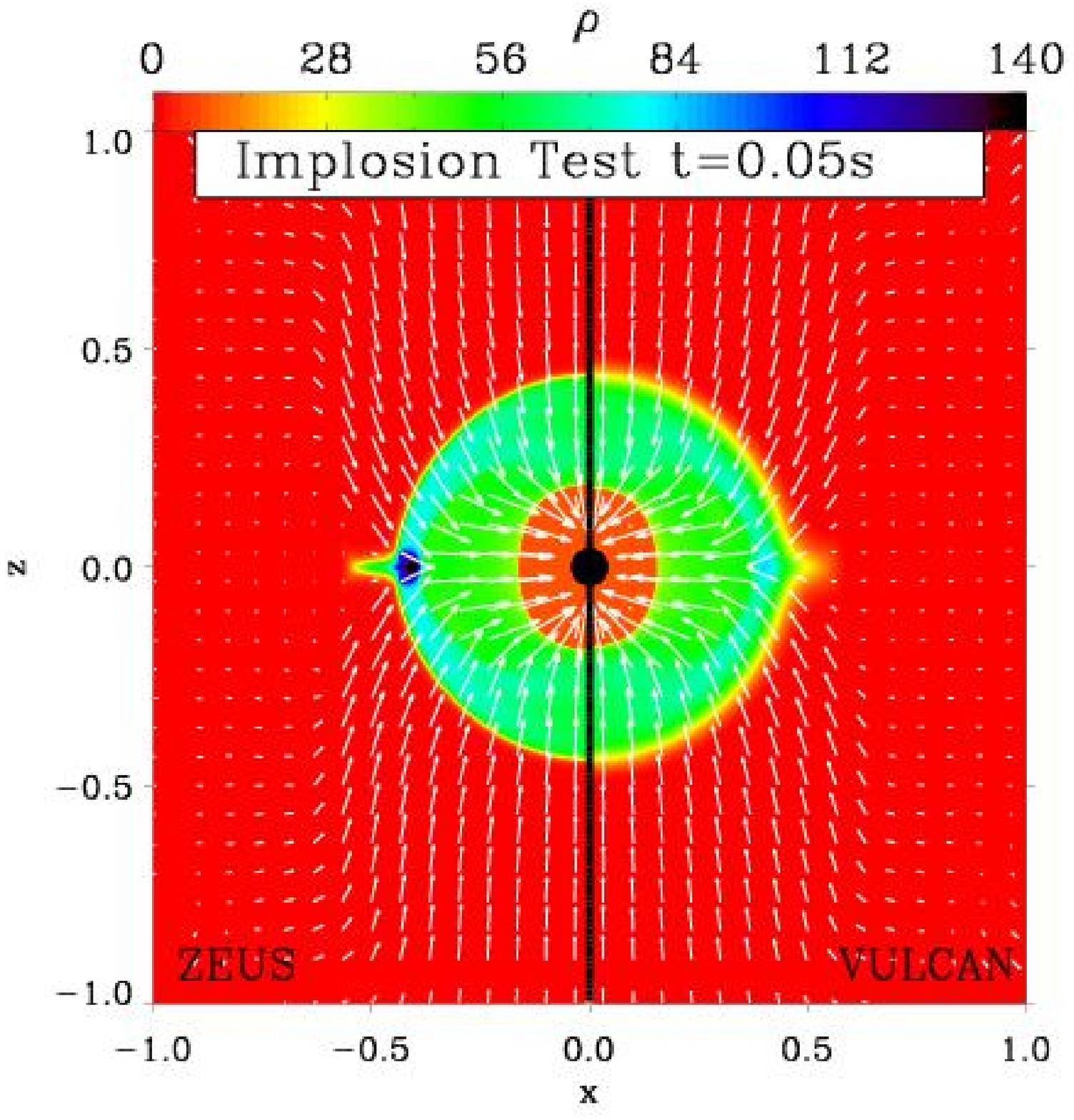}{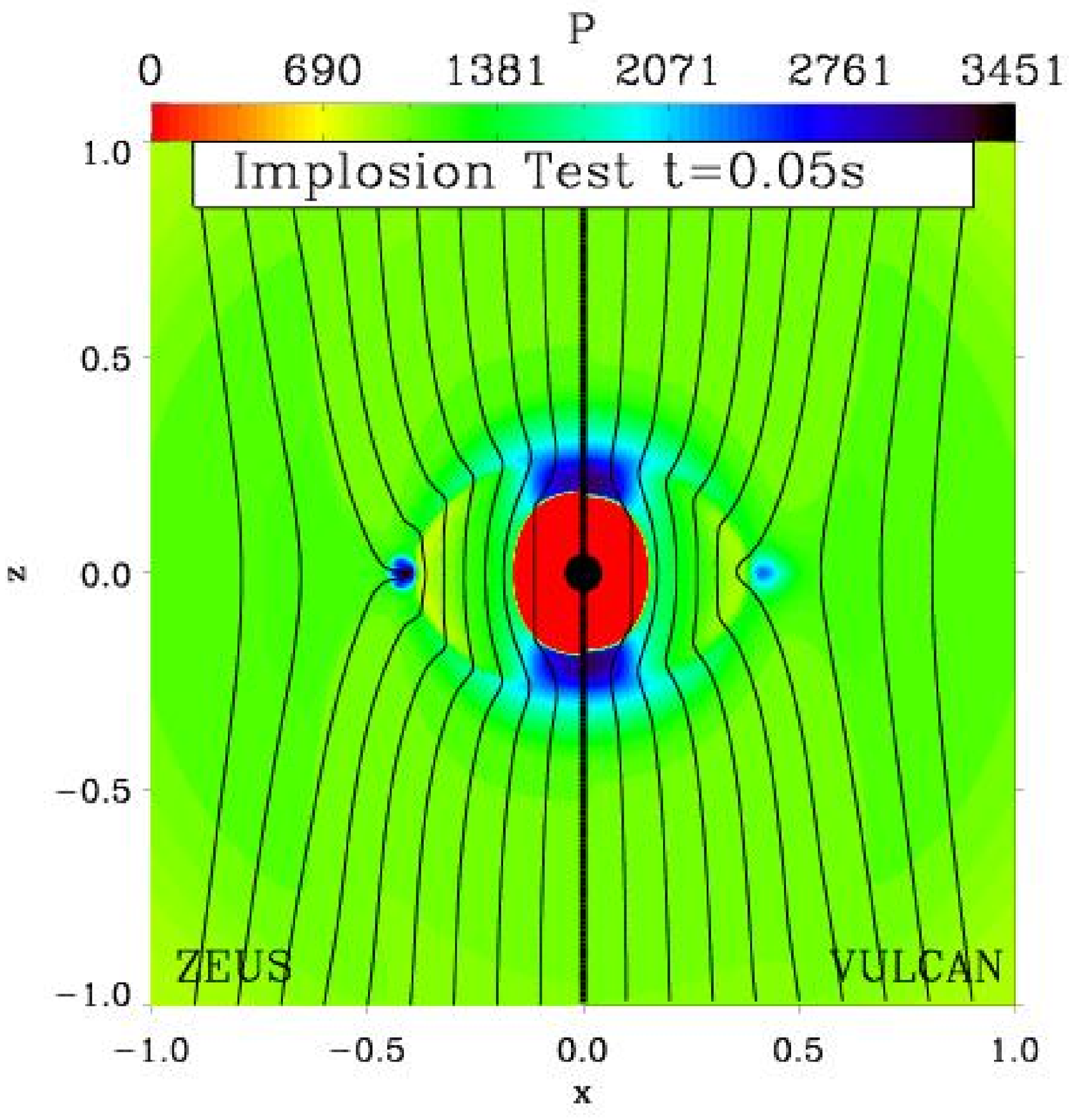}
\plottwo{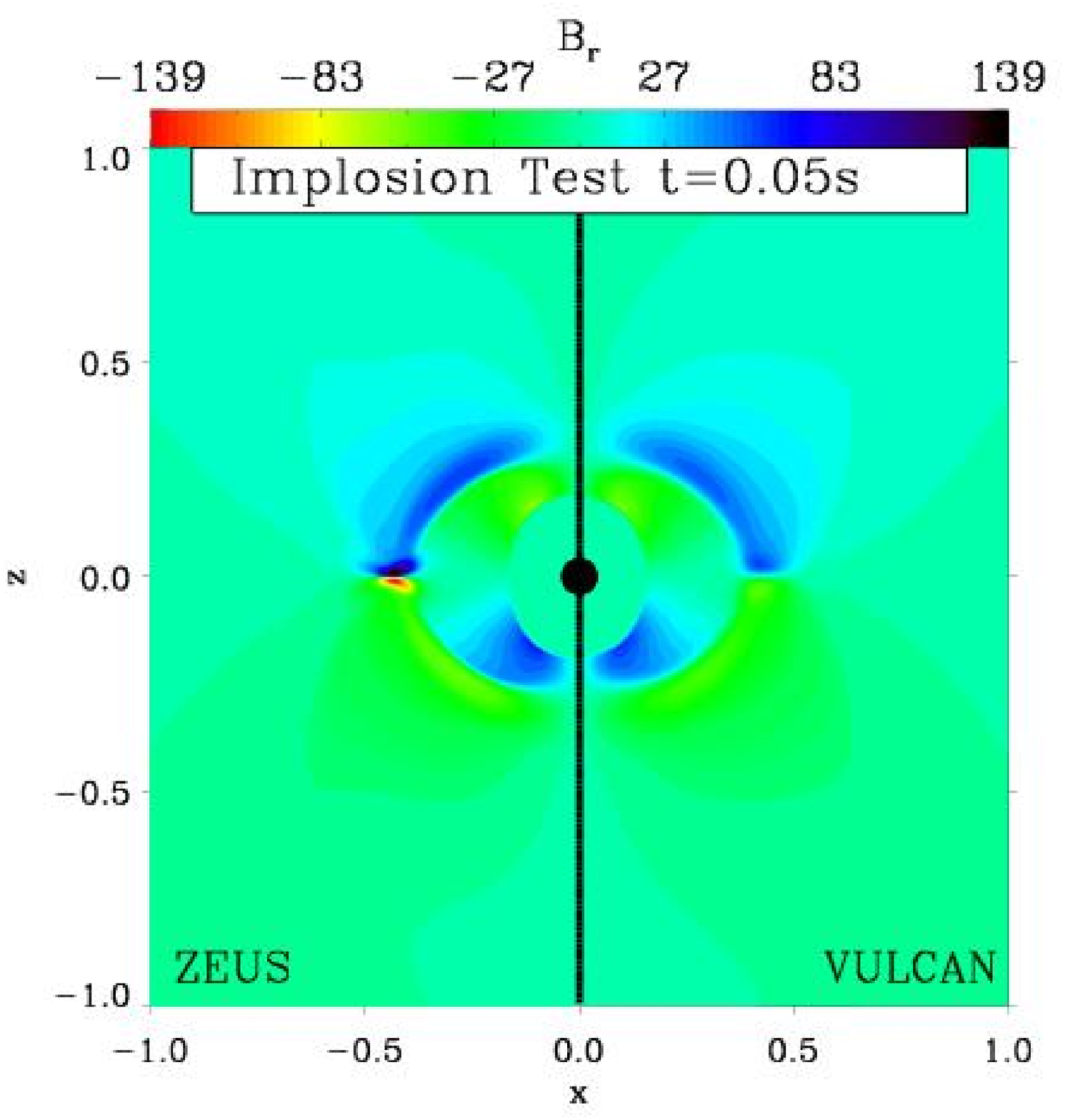}{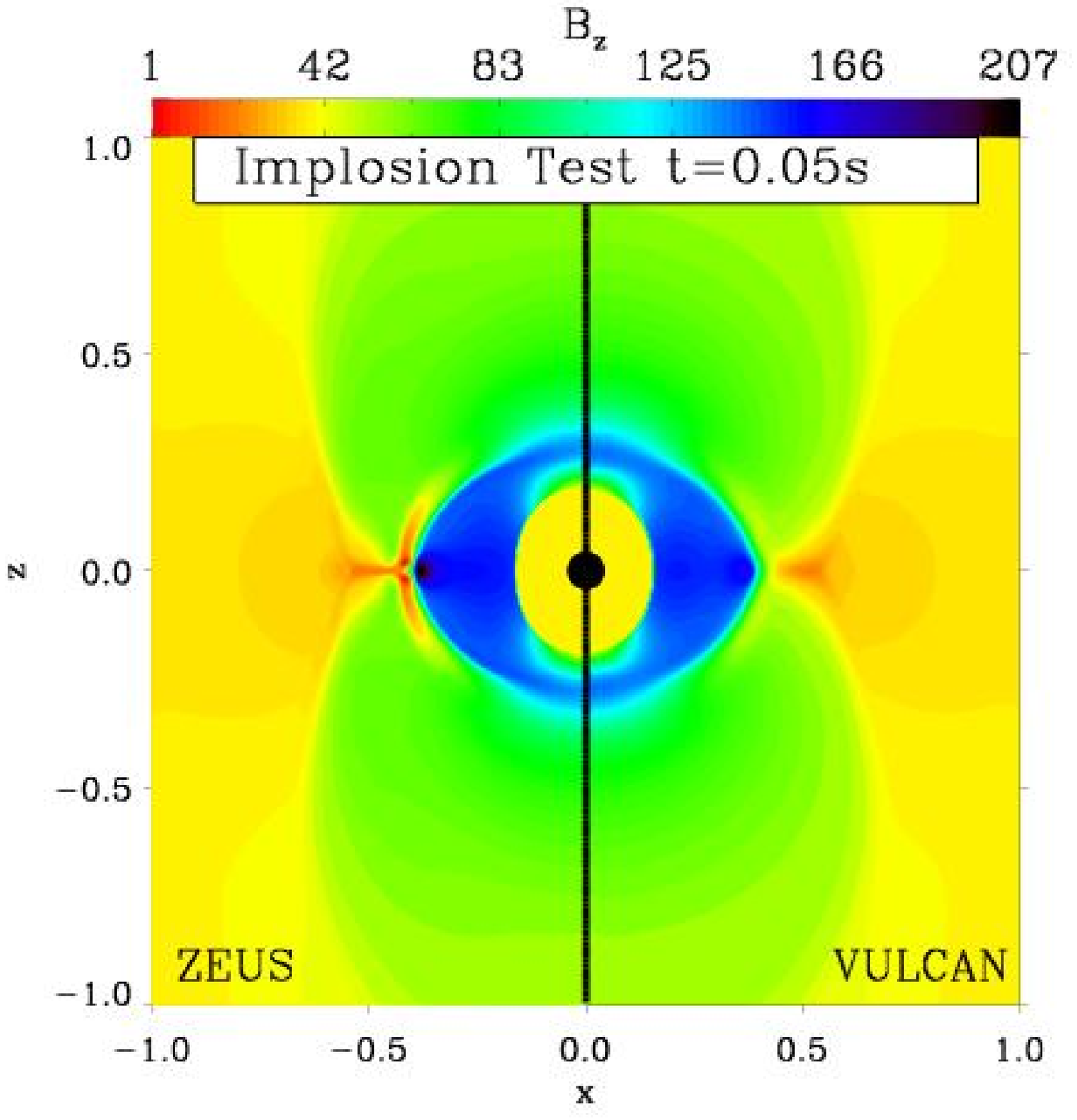}
\caption{Colormaps of the high-resolution non-rotating implosion test at t=0.05, just before the shock
 bounces at the inner boundary:
 density (top left), pressure (top right), $B_r$ (bottom left) and $B_z$ (bottom right).
 Each panel shows the ZEUS/2D (left) and VULCAN/2D (right) results.
 Velocity vectors are overplotted in white in the top-left box, with a maximum length set to 10\% of the width
 of the panel, and a maximum magnitude of 15 (all quantities are dimensionless).
 Magnetic field lines are overplotted in black in the top-right box. 
 Note the top-bottom symmetry of the z-component of the magnetic field, and the anti-symmetry of the r-component,
 best rendered by the morphology of the poloidal field lines (top right panel).
}
\label{fig_imp_norot_t0pt05}
\end{figure*}

\clearpage

\begin{figure*}[tp!]
\epsscale{1.1}
\plottwo{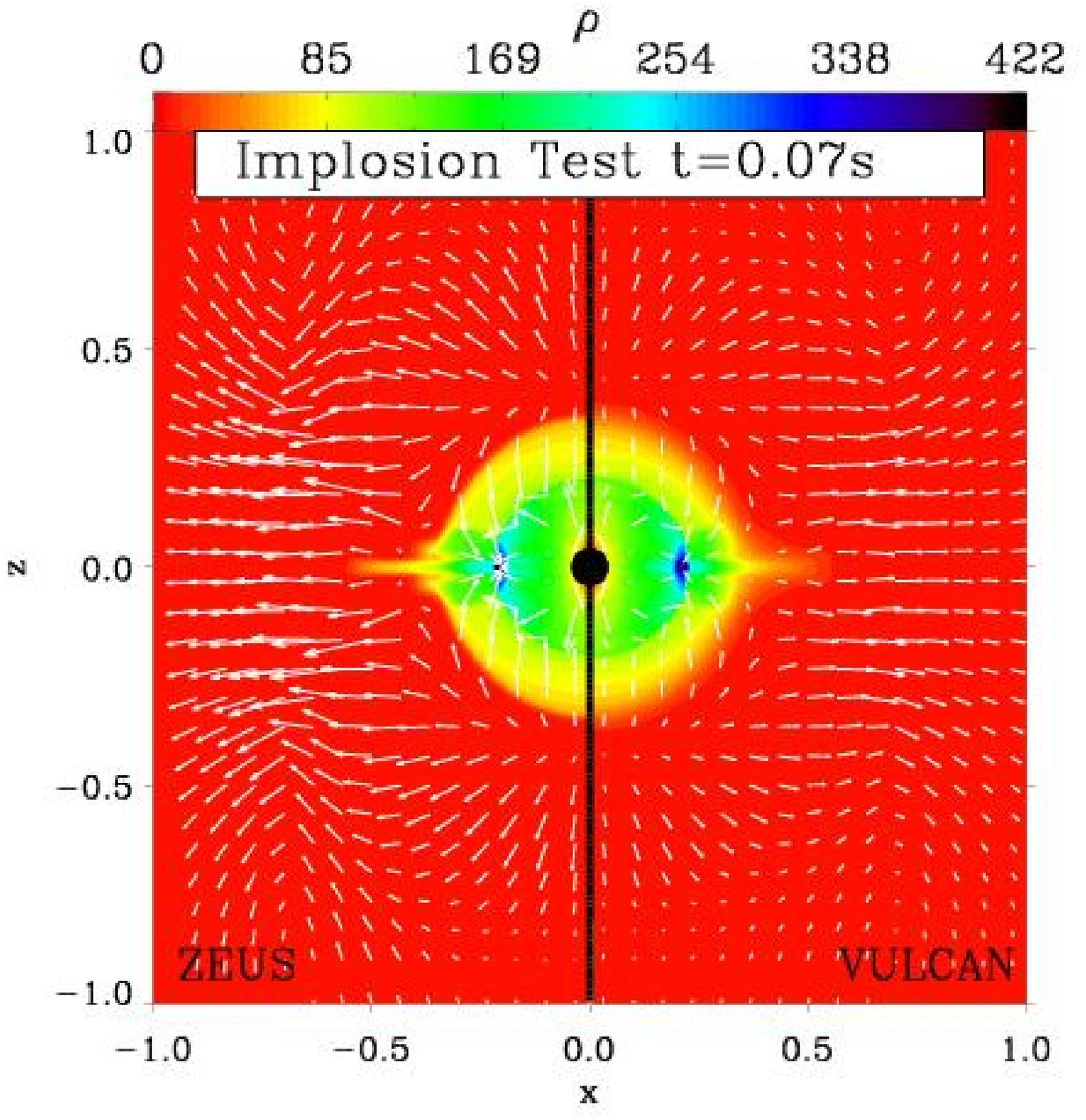}{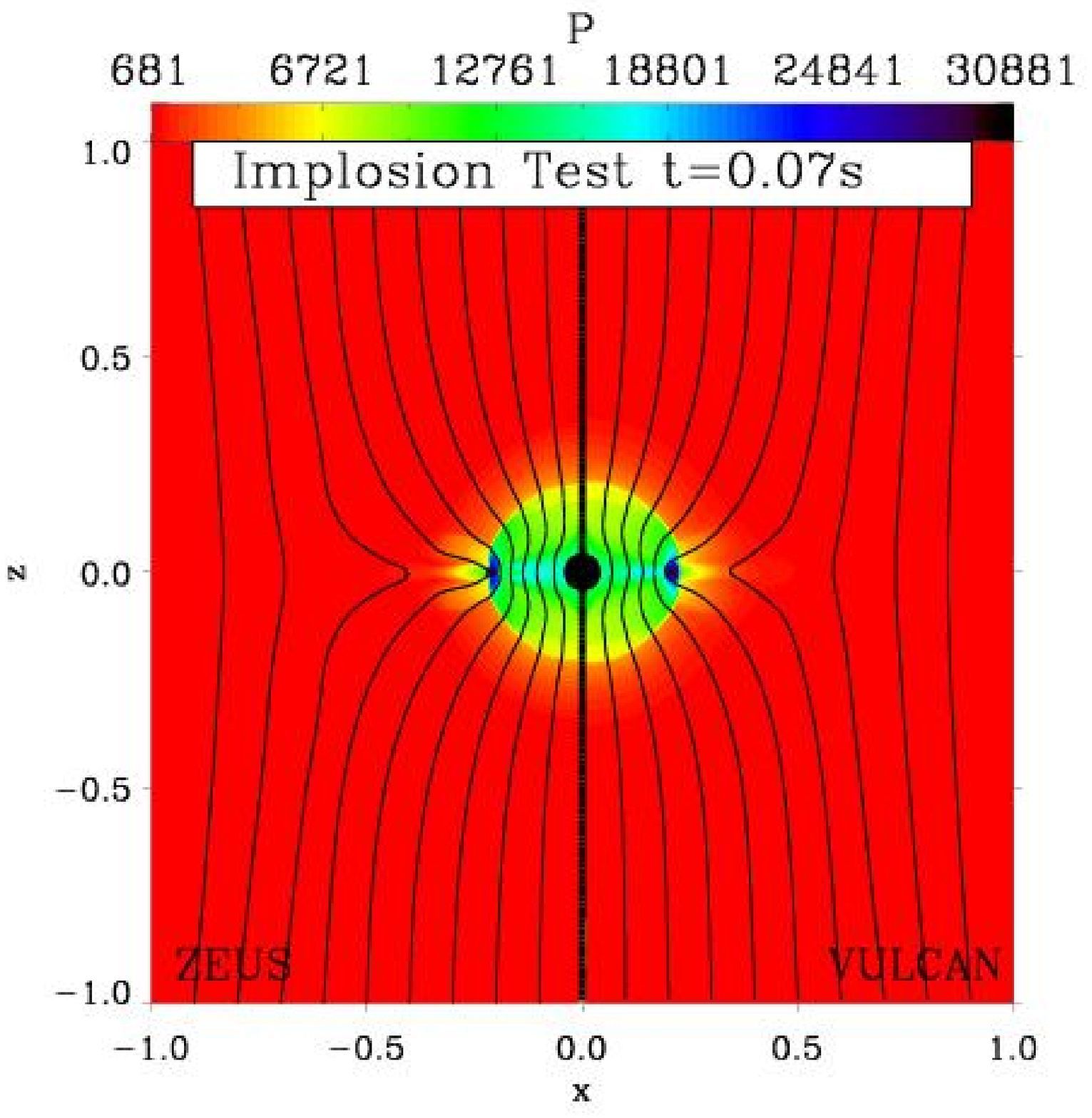}
\plottwo{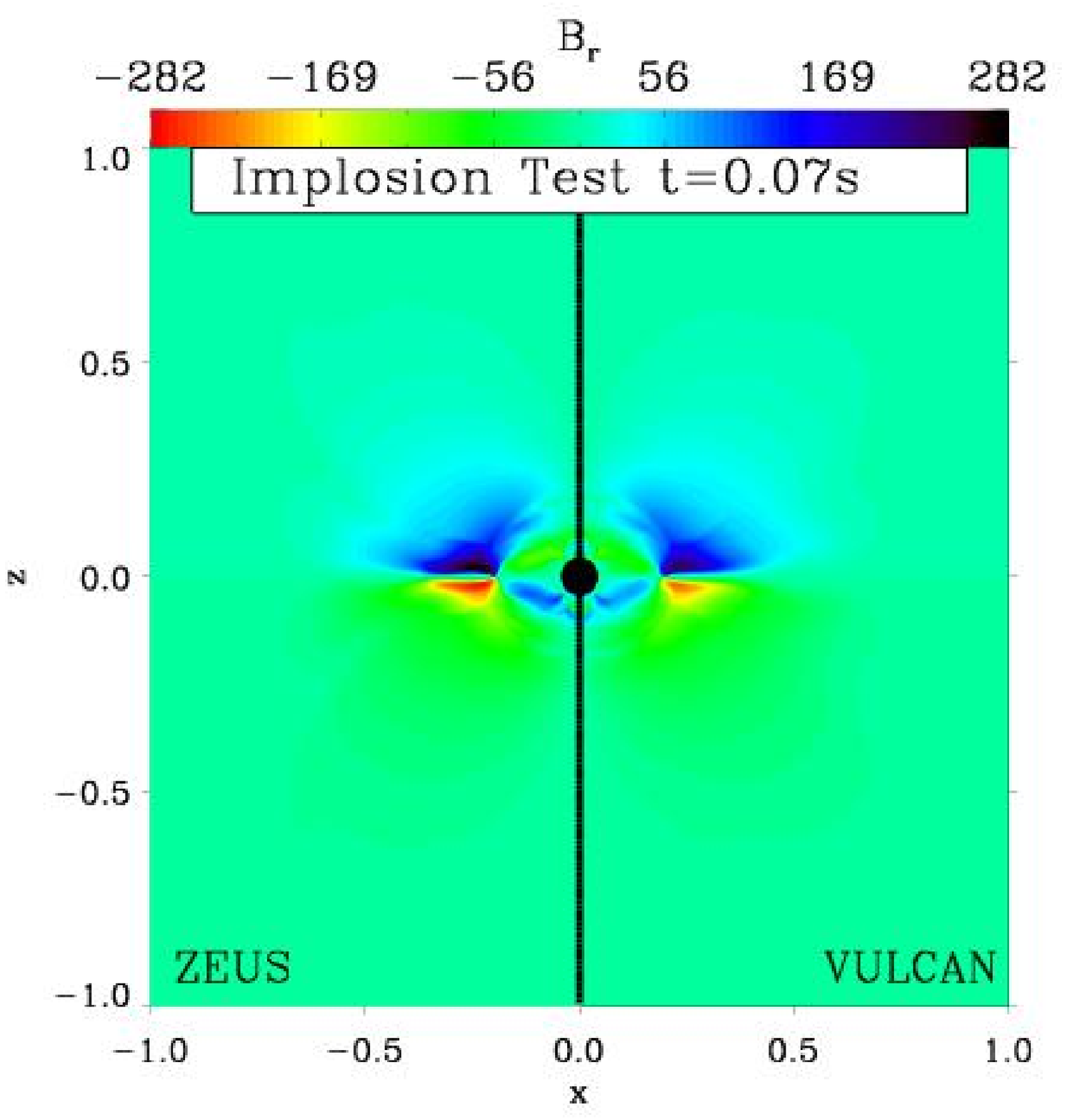}{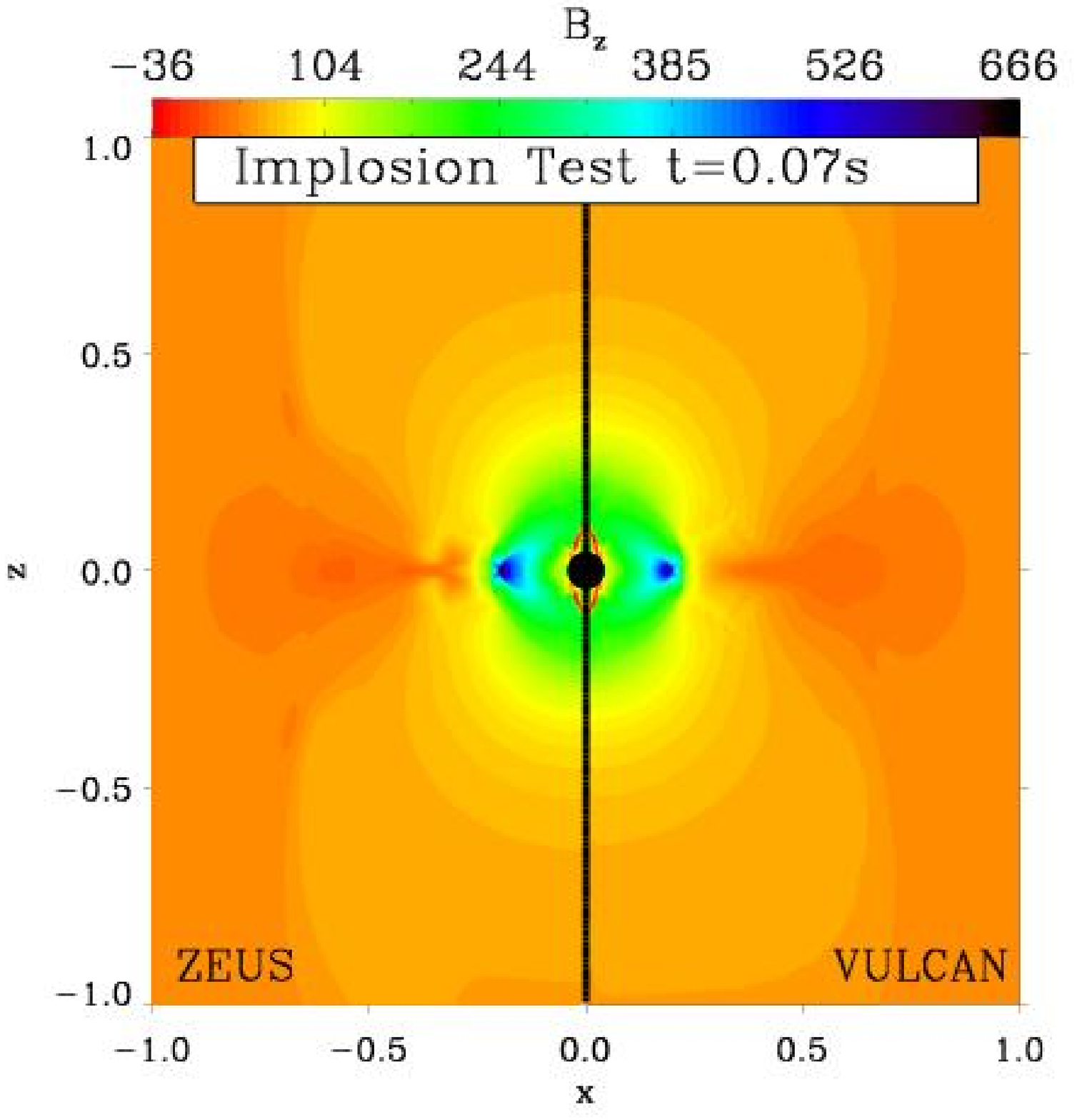}
\caption{Same as Fig.~\ref{fig_imp_norot_t0pt05}, but at time $t=0.07$, corresponding to
a time after bounce of $\sim$0.01 seconds.  
}
\label{fig_imp_norot_t0pt07}
\end{figure*}

\clearpage

\begin{figure*}[tp!]
\epsscale{1.1}
\plottwo{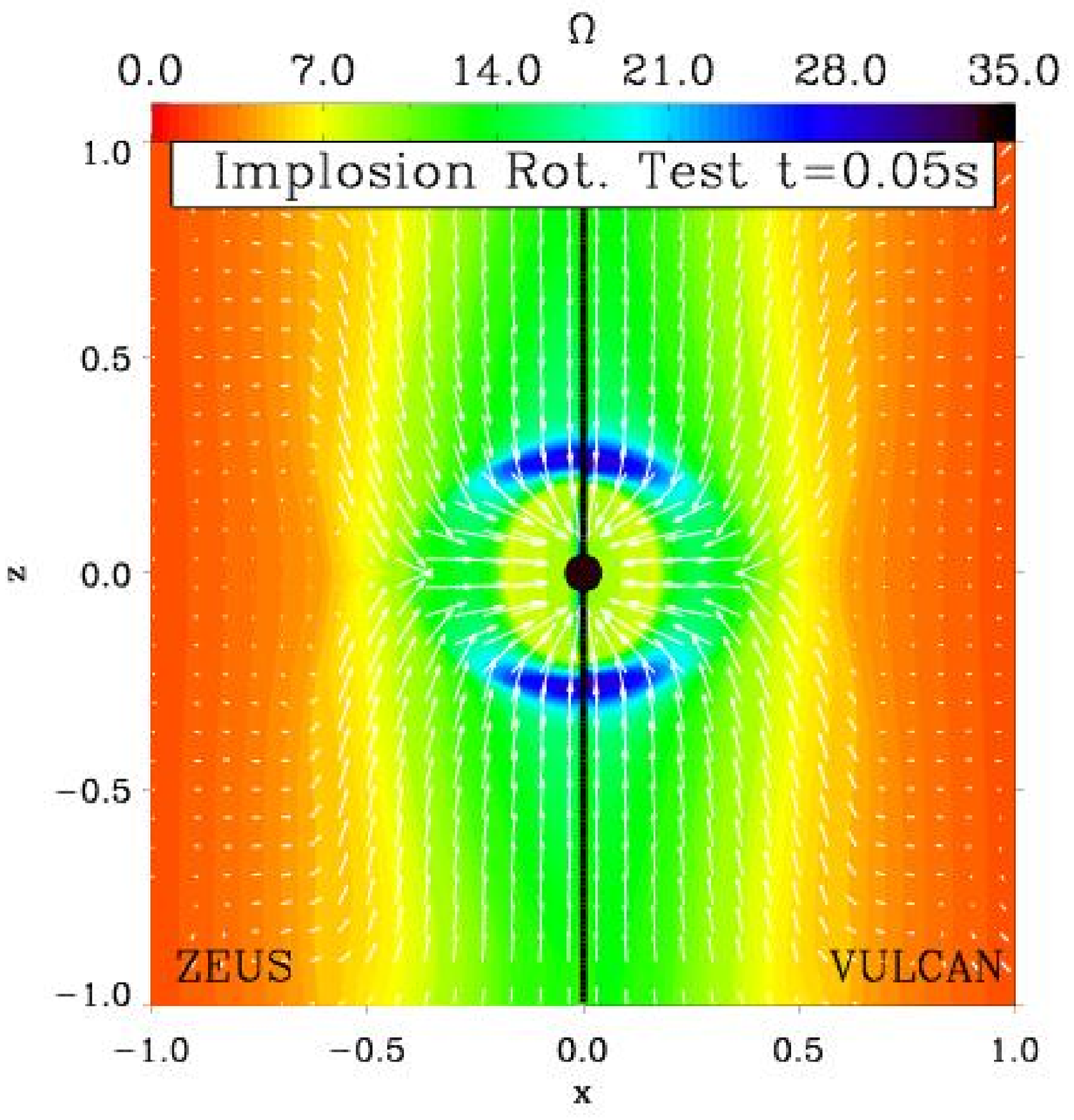}{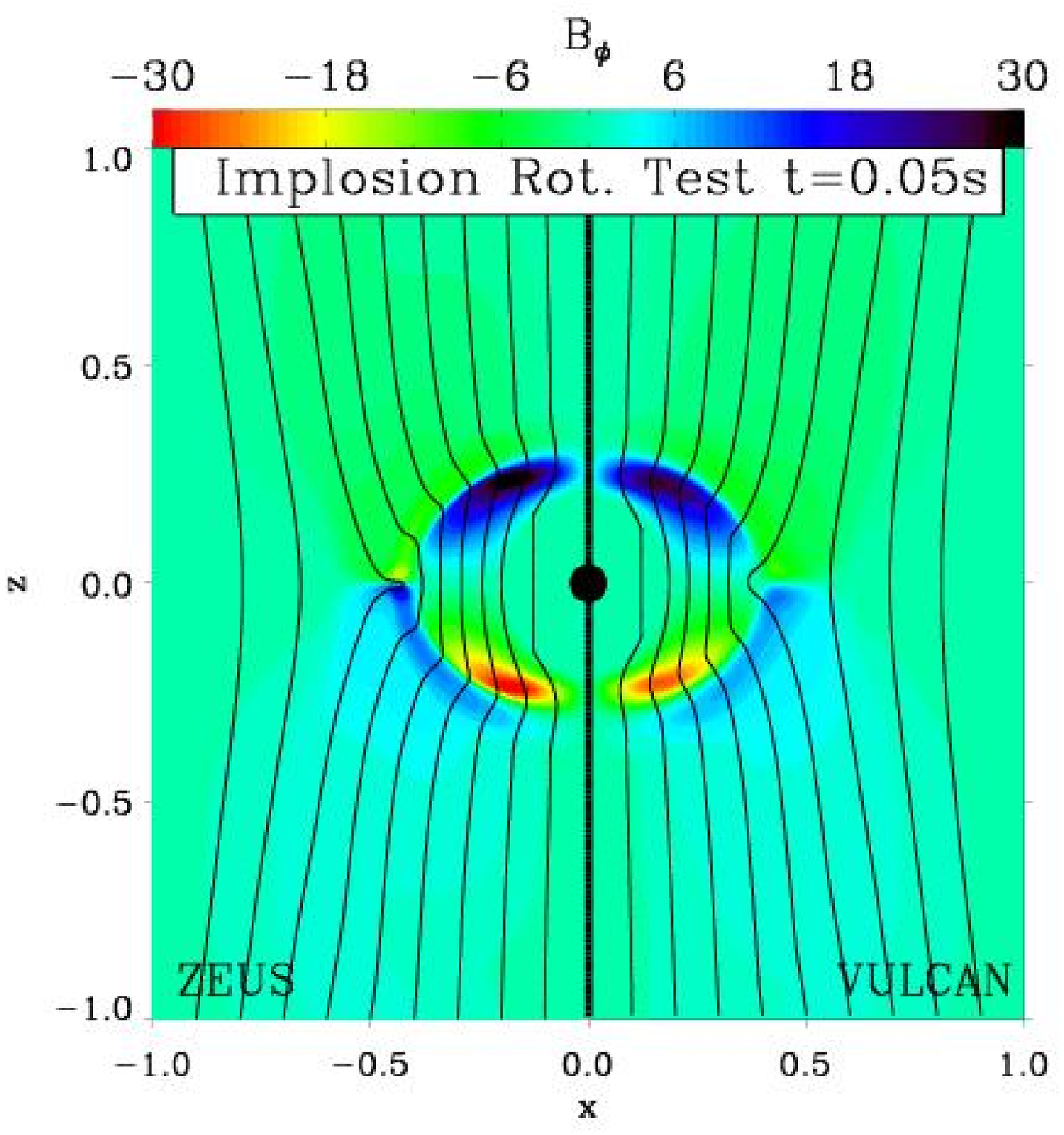}
\plottwo{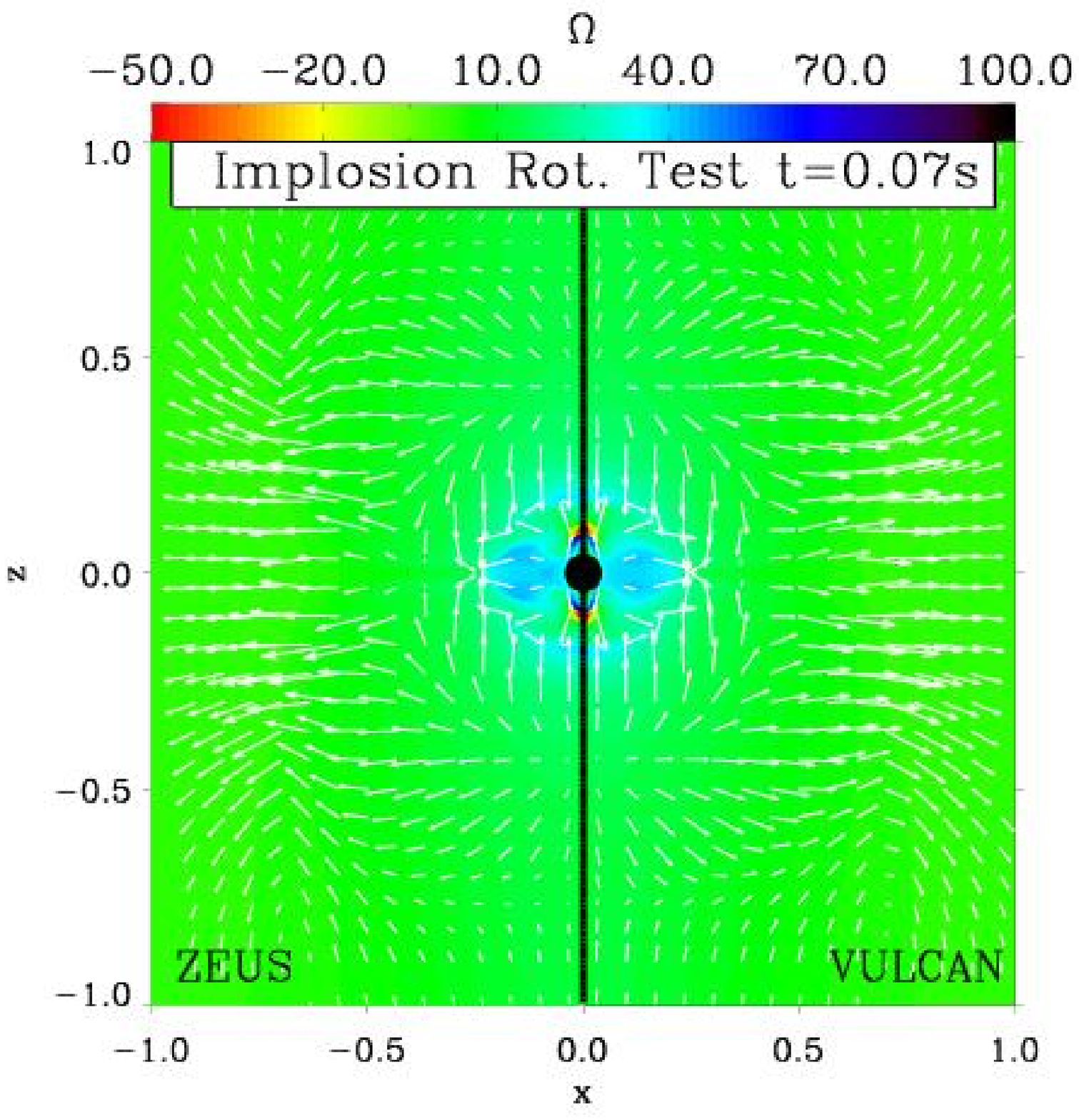}{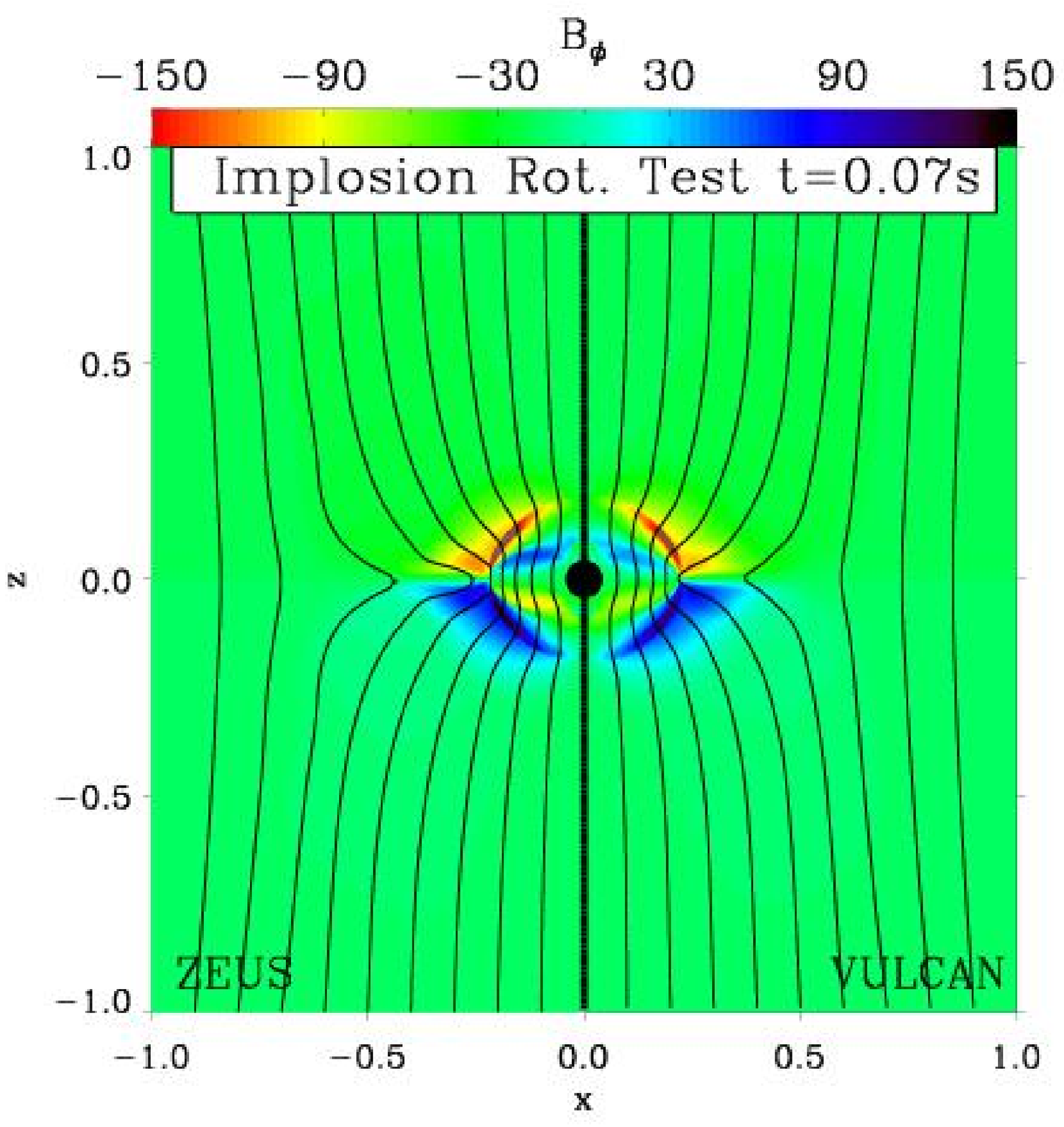}
\caption{Colormaps of the rotating implosion test at t=0.05 and t=0.07:
The angular velocity $\Omega$ (left column), the toroidal magnetic field
$B_{\phi}$ (right column), at times $t=0.05$ (before bounce; top row) 
and $t=0.07$ (after bounce; bottom row).
Each panel shows the ZEUS/2D (left) and VULCAN/2D (right) results.
Again, note the top-bottom symmetry of the z-component of the 
magnetic field and the anti-symmetry of the r-component,
best rendered by the morphology of the poloidal field lines (right panels).
}
\label{fig_imp_rot}
\end{figure*}

\clearpage

\begin{figure*}[tp!]
\epsscale{1.1}
\plottwo{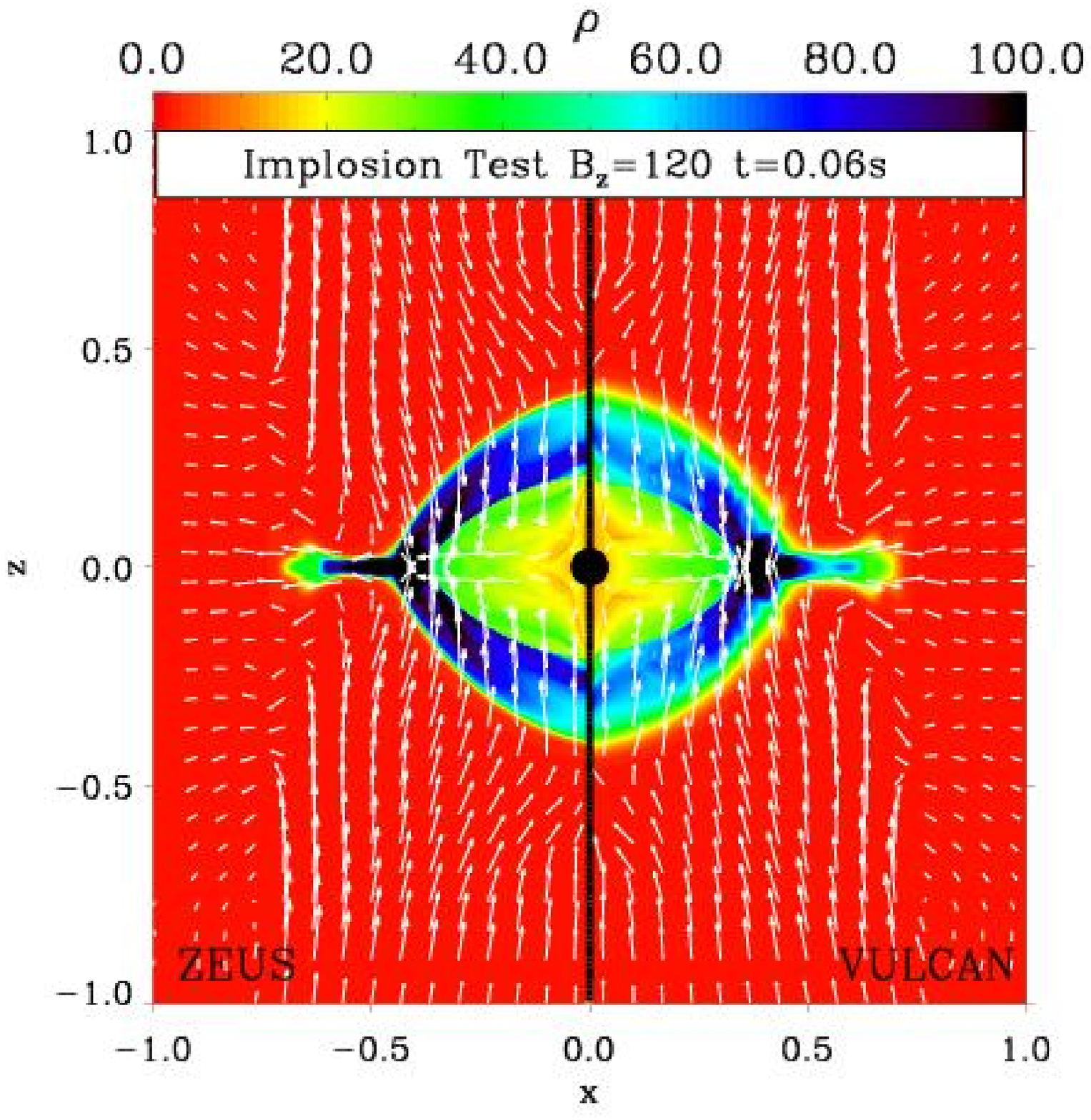}{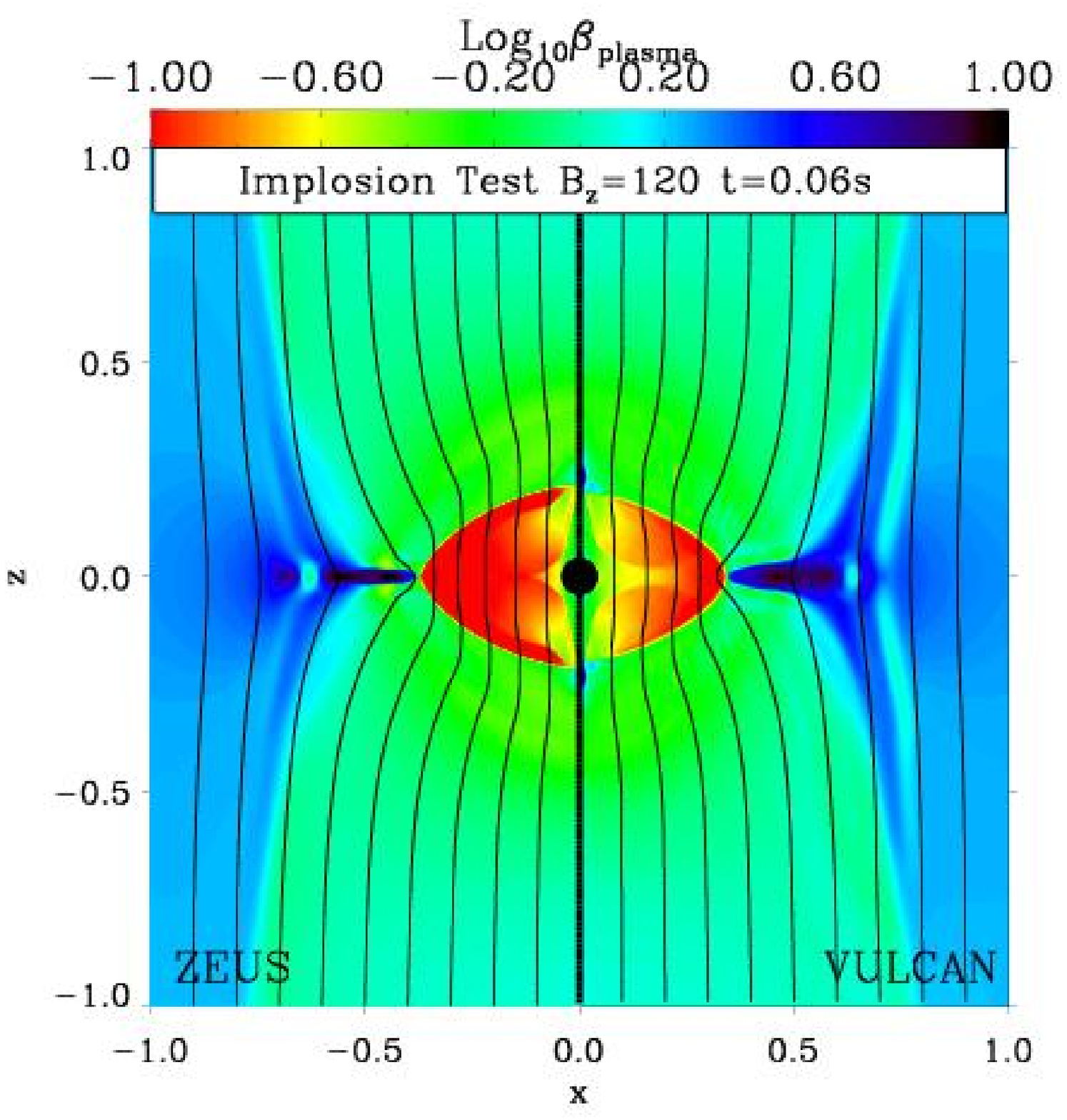}
\plottwo{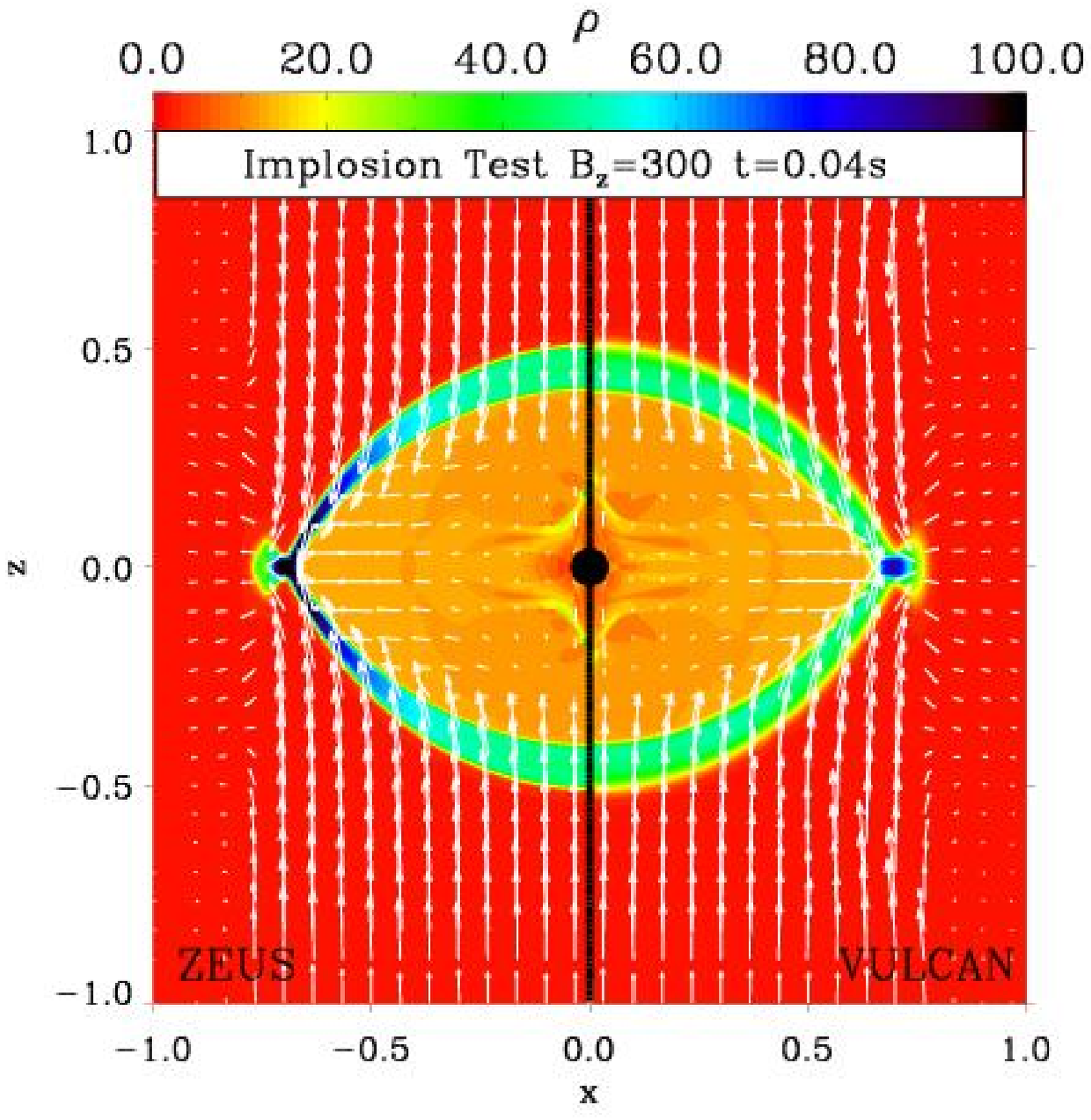}{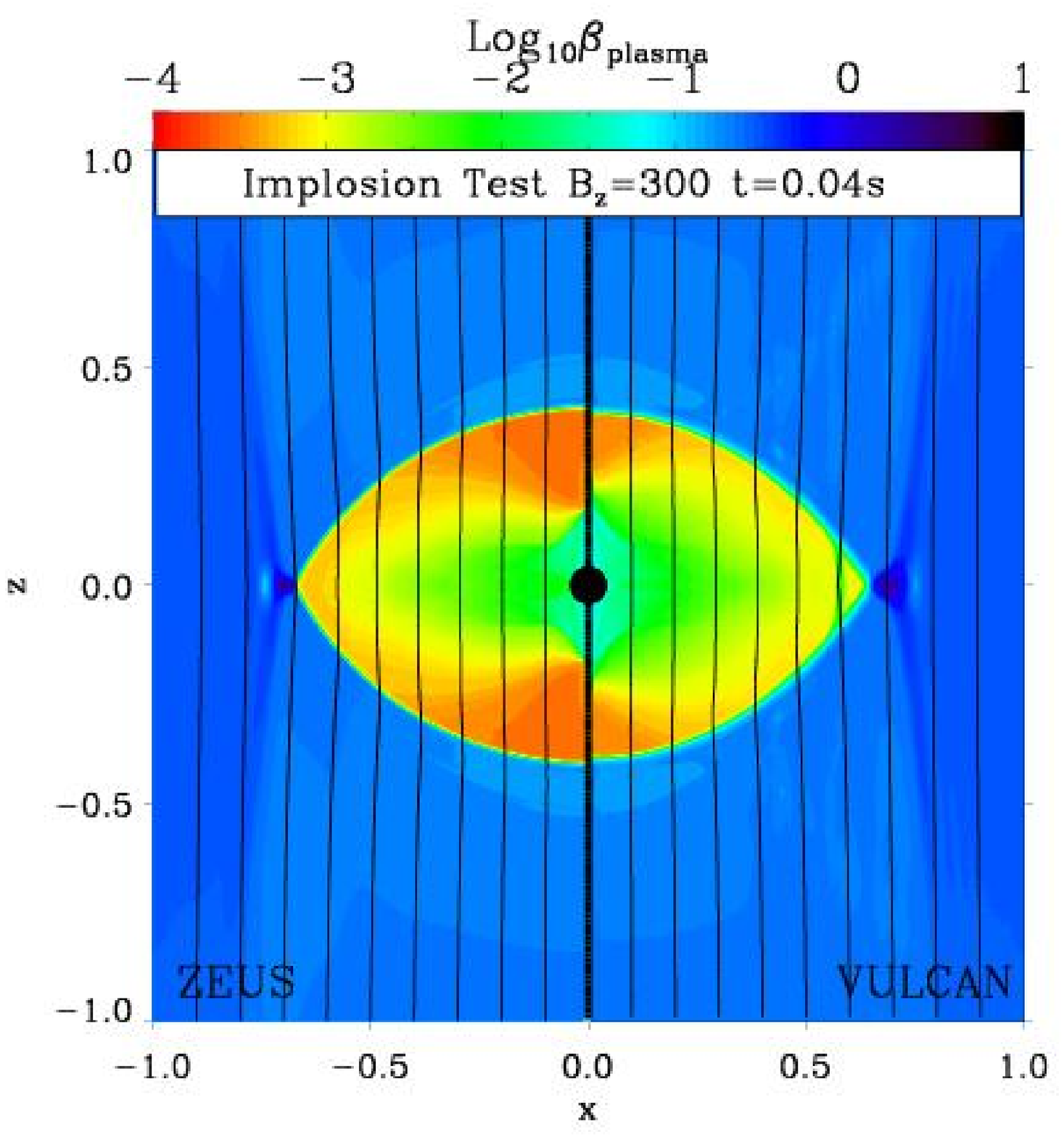}
\caption{Colormaps of the high-resolution non-rotating implosion
 tests with dominant initial magnetic fields 
 ($B_z=120$, $t=0.06$, top row; $B_z=300$, $t=0.04$, bottom row).
  Shown are the density (left column)  and the quantity 
  $\beta_{\rm plasma} = P / (B^2/8\pi)$ (right column).
  Each panel shows results for the ZEUS/2D (left) and VULCAN/2D (right).
Again, note the top-bottom symmetry of the z-component of the
magnetic field and the anti-symmetry of the r-component,
best rendered by the morphology of the poloidal field lines (right panels).
}
\label{fig_imp_highB}
\end{figure*}


\begin{thebibliography}

\bibitem[Akiyama et al.(2003)]{akiyama:03}
Akiyama, S., Wheeler, J.C., Meier, D.L., \& Lichtenstadt, I. 2003,
\apj, 584, 954

\bibitem[Akiyama \& Wheeler(2005)]{akiyama:05}
Akiyama, S. \& Wheeler, J.C. 2005, \apj, 629, 414  

\bibitem[Aloy et al.(1999a)]{aloya} Aloy, M.A., Iba\'nez, J.M., Marti, J.M., 
G\'omez, J.-L., \& M\"uller, E. 1999a, \apj, 523, L125

\bibitem[Aloy et al.(1999b)]{aloyb} Aloy, M.A., Iba\'nez, J.M., Marti, J.M., \& M\"uller, E. 1999b,
\apjs, 122, 151

\bibitem[Anninos,~Fragile,~\&~Salmonson(2005)]{anninos} Anninos, P., Fragile, P.C., 
\& Salmonson, J.D. 2005, \apj, 635, 723

\bibitem[Ardeljan et al.(2000)]{ABM1} Ardeljan, N.V., Bisnovatyi-Kogan, 
G.S., \& Moiseenko, S.G. 2000, \aap, 355, 1181,

\bibitem[Ardeljan et al.(2005)]{ABM2} Ardeljan, N.V., Bisnovatyi-Kogan, 
G.S., \& Moiseenko, S.G. 2005, \mnras, 359, 333

\bibitem[Balbus \& Hawley(1991)]{balbus} Balbus, S.A. \& Hawley, J.F. 1991, \apj, 376, 222

\bibitem[Balsara~\&~Kim(2004)]{balsara} Balsara, D.S. \& Kim, J. 2004, \apj, 602, 1079

\bibitem[Bisnovatyi-Kogan et al.(1976)]{BPS} Bisnovatyi-Kogan, 
G.S., Popov, I.P., \& Samokhin, A.A. 1976, \apss, 41, 287

\bibitem[Blondin et al.(2003)]{sasi2} Blondin, J.M., Mezzacappa, 
A., \& DeMarino, C. 2003, \apj, 584, 971

\bibitem[Brio \& Wu(1988)]{BW} Brio, M., \& Wu, C.C. 1988, J. Comp. Phys., 75, 400

\bibitem[Bucciantini et al.(2006)]{bucciantini} Bucciantini, N., Thompson,
T.A., Arons, J., Quataert, E., \& Del Zanna, L. 2006,
\mnras, 368, 1717

\bibitem[Burrows et al.(2006a)]{new_mech} Burrows, A., Livne, E., 
Dessart, L., Ott, C.D., \& Murphy, J. 2006, \apj, 640, 878

\bibitem[Burrows et al.(2006b)]{new_mechb} Burrows, A., Livne, E., 
Dessart, L., Ott, C.D., \& Murphy, J. 2006, \apj, 640, 878

\bibitem[Calder \etal 2002]{cal02} Calder, A.C. et al. 2002, \apjs, 143, 201

\bibitem[Chandrasekhar(1961)]{Ch} Chandrasekhar, S. 1961, in ``Hydrodynamic and Hydromagnetic Stability'', 
Oxford Univ. Press

\bibitem[Colella(1990)]{colella} Colella, P. 1990, J. Comp. Phys., 87, 171

\bibitem[Courant \& Friedrichs(1948)]{CFL} Courant, R., \& 
Friedrichs, K.O. 1948, in ``Pure and Applied Mathematics'', New York: 
Interscience

\bibitem[Dai \& Woodward(1994)]{DW94} Dai, W. \& Woodward, P.R. 1994, J. Comp. Phys., 111, 354

\bibitem[Dai \& Woodward(1998)]{DW98} Dai, W. \& Woodward, P.R. 1998, \apj,494,317

\bibitem[Dessart et al. 2006a]{dessarta} Dessart, L., Burrows, A., Ott, C.D., Livne, E.,
Yoon, S.-Y., \& Langer, N. 2006a, \apj, 644, 1063

\bibitem[Dessart et al. 2006b]{dessartb} Dessart, L., Burrows, A., Livne, E., \& Ott, C.D.
2006b, \apj, 645, 534  

\bibitem[Dimonte \etal 2004]{di04} Dimonte, G. et al. 2004, Phys. of Fluids, 16, 1668

\bibitem[Evans \& Hawley 1988]{EH88} Evans, C.R. \& Hawley, J.F. 1988, \apj, 332, 659

\bibitem[Etienne, Liu, \& Shapiro(2006)]{etienne}  Etienne, Z.B., Liu, Y.T.,
\& Shapiro, S.L. 2006, \prd, 74, 044030

\bibitem[Foglizzo et al.(2005)]{sasi1} Foglizzo, T., Galleti, P. \& Ruffert, M. 2005, A\&A, 435, 397

\bibitem[Frenk et al. 1999]{Fr99} Frenk, C.S. et al. 1999, \apj, 525, 554

\bibitem[Gardiner \& Stone(2005)]{gardiner} Gardiner, T.A. \& Stone, J.M. 
2005, J. Comp. Phys., 205, 509 

\bibitem[Heger et al.(2004)]{heg} Heger, A., Woosley, 
S.~E., Langer, N., \& Spruit, H.~C. 2004, IAU Symposium, 215, 591 

\bibitem[Heger,~Woosley,~\&~Spruit(2005)]{heger05} Heger, A., Woosley,
S.E., \& Spruit, H. 2005, \apj, 626, 350 (astro-ph/0409422)

\bibitem[Holmes \etal 1999]{Ho99} Holmes, R.L., Dimonte, G., Fryxell, B.A., Gittings, M.L., 
  Grove, J.W., Schneider, M., Sharp, D.H., Velikovich, A.L., Weaver, R.P., \& 
  Zhang, Q. 1999, J. Fluid Mech., 389, 55

\bibitem[Kotake et al.(2004)]{kot04} Kotake, K., Sawai, H., Yamada, S., \& Sato, K. 2004, \apj, 608, 391

\bibitem[Landau \& Lifshitz(1960)]{LL} Landau, L.D. \& Lifshitz, E.M. 1960, in
``Electrodynamics of Continuous Media,'' Pergamon Press

\bibitem[LeBlanc \& Wilson(1970)]{LBW} LeBlanc, J.M., \& Wilson, J.R. 1970, \apj, 161, 541

\bibitem[Livne(1993)]{V2D} Livne, E. 1993, \apj, 412, 634

\bibitem[Livne et al.(2004)]{livne04}Livne, E., Burrows, A., Walder, R.,
Thompson, T.A., and Lichtenstadt, I. 2004, \apj, 609, 277

\bibitem[Masada et al.(2006)]{masada} Masada, Y., Sano, T., \& Shibata, K. 2006, accepted to \apj (astro-ph/0610023)

\bibitem[MacFadyen \& Woosley(1999)]{macfadyen} MacFadyen, A.I. \& Woosley, S.E. 1999, \apj, 524, 262

\bibitem[Meier \etal 1976]{Me76} Meier, D.L., Epstein, R.I., Arnett, W.D \& Schramm, D.N., 1976,\apj,204,869

\bibitem[Metzger, Thompson, \& Quataert(2006)]{metzger} Metzger, B., Thompson, T.A., \&
Quataert, E. 2006, astro-ph/0608682

\bibitem[Mizuno et al.(2004)]{mizuno} Mizuno, Y., Yamada, S., Koide, S., \& Shibata, K. 2004, \apj, 606, 395

\bibitem[Moiseenko et al.(2006)]{moiseenko} Moiseenko, S.G.,
Bisnovatyi-Kogan, G.S., \& Ardeljan, N.V. 2006, \mnras, 370, 501

\bibitem[Noble et al.(2006)]{noble} Noble, S.C., Gammie, C.F., McKinney,
J.C., \& Del Zanna, L. 2006, \apj, 641, 626

\bibitem[Ohnishi, Kotake, \& Yamada(2005)]{ohnishi} Ohnishi, N., Kotake, K., \& Yamada, S. 2005, astro-ph/0509765

\bibitem[Obergaulinger et al.(2006)]{obergaulinger} Obergaulinger,
M., Aloy, M.A., \& M\"uller, E.\ 2006, \aap, 450, 1107

\bibitem[Ott et al.(2006a)]{ott06a} Ott, C.D., Burrows, A., Dessart, L., \& Livne, E. 2006a,
\apj~Suppl., 164, 130 

\bibitem[Ott et al. 2006b]{ott06b} Ott, C.D., Burrows, A., Dessart, L., \& Livne, E. 2006b,
\prl,  96, 201102  

\bibitem[Pessah \& Psaltis(2005)]{2005ApJ...628..879P} Pessah, M.E., \&
 Psaltis, D. 2005, \apj, 628, 879
 
\bibitem[Pessah, Chan, \& Psaltis 2006]{pessah} Pessah, M.E., Chan, C.,
 \& Psaltis, D. 2006, \mnras, in press (astro-ph/0603178)

\bibitem[Price~\&~Monaghan(2005)]{monaghan} Price, D.J. \& Monaghan, J.J. 2005, \mnras, 364, 384

\bibitem[Price~\&~Rosswog(2006)]{priceross} Price, D.J. \& Rosswog, S. 2006, Science, 312, 5774

\bibitem[Proga(2005)]{proga} Proga, D. 2005, \apj, 629, 397

\bibitem[Ryu \& Jones(1995)]{RJ95} Ryu, D., \& Jones, T.W. 1995, \apj, 442, 228


\bibitem[Shibata~\&~Sekiguchi(2005)]{shibata} Shibata, M. \& Sekiguchi, Y. 2005, \prd, 72, 044014

\bibitem[Stone \& Norman(1992a)]{Zeus} Stone, J.M., \& Norman, M.L. 1992a, \apjs, 80, 753 

\bibitem[Stone \& Norman(1992b)]{Zeusmhd} Stone, J.M., \& Norman, M.L. 1992, \apjs, 80, 791 

\bibitem[Stone et al.(1992)]{S92} Stone, J.M., Hawley, J.F., Evans, C.R., \& Norman, M.L. 1992, \apj, 388, 415

\bibitem[Symbalisty(1984)]{symbalisty} Symbalisty, E.M.D. 1984, \apj, 285, 729

\bibitem[Thompson,~Chang,~\&~Quataert(2004)]{tcq} Thompson, T.A., Chang, P., \& Quataert, E. 2004,
\apj, 611, 393

\bibitem[Thompson,~Quataert,~\&~Burrows(2005)]{tqb} Thompson, T.A., Quataert,
E., \& Burrows, A. 2005,
\apj, 620, 861

\bibitem[Toth (2000)]{TH} Toth, G. 2000, \jcp, 161, 605

\bibitem[Uzdensky \& MacFadyen(2006a)]{uzdensky} Uzdensky, D.A. \& MacFadyen, A.I. 2006a, \apj, 647, 1192

\bibitem[Uzdensky \& MacFadyen(2006b)]{uzdenskyb} Uzdensky, D.A. \& MacFadyen, A.I. 2006b, astro-ph/0609047

\bibitem[Wilson, Mathews, \& Dalhed(2005)]{dalhed} Wilson, J.R., Mathews, G.J., \& Dalhed, H.E. 2005, \apj, 628, 335

\bibitem[Walder et al. 2005]{walder} Walder, R., Burrows, A., Ott,
C.D., Livne, E., Lichtenstadt, I., \& Jarrah, M. 2005, \apj, 626, 317

\bibitem[Yamada \& Sawai(2004)]{y04} Yamada, S., \& Sawai, H. 2004, \apj, 608, 907 

\bibitem[van Leer(1979)]{vanLeer} van Leer, B. 1979, J. Comp. Phys., 32, 101

\bibitem[Ziegler,~Dolag,~\&~Bartelmann(2006)]{ziegler} Ziegler, E., Dolag, K., \& Bartelmann, M. 2006,
Astronomische Nachrichten, 327, 607 

\end{thebibliography}
\end{document}